\documentclass[10pt,letterpaper]{article}
\usepackage[top=0.85in,left=2.75in,footskip=0.75in,marginparwidth=2in]{geometry}

\usepackage[utf8]{inputenc}

\usepackage{cite}

\usepackage{nameref,hyperref}

\usepackage[right]{lineno}

\usepackage{float}

\usepackage{amsmath}

\usepackage{microtype}
\DisableLigatures[f]{encoding = *, family = * }

\raggedright
\usepackage{changepage}

\usepackage[aboveskip=1pt,labelfont=bf,labelsep=period,singlelinecheck=off]{caption}

\makeatletter
\renewcommand{\@biblabel}[1]{\quad#1.}
\makeatother

\usepackage{lastpage,fancyhdr,graphicx}
\usepackage{epstopdf}
\fancyheadoffset[L]{2.25in}
\fancyfootoffset[L]{2.25in}

\usepackage{color}

\definecolor{Gray}{gray}{.25}

\usepackage{graphicx}

\usepackage{sidecap}

\usepackage{wrapfig}
\usepackage[pscoord]{eso-pic}
\usepackage[fulladjust]{marginnote}
\reversemarginpar

\begin{document}
\vspace*{0.35in}

% title goes here:
\begin{flushleft}
{\Large
\textbf\newline{An analytical framework for understanding tropical Meridional Modes}
}
\newline
% authors go here:
\\
Cristian Martinez-Villalobos\textsuperscript{1,*},
Daniel J. Vimont\textsuperscript{1}
\\
\bigskip
\textbf{Atmospheric and Oceanic Sciences Department, University of Wisconsin -- Madison, Madison, WI, U.S.A.}
\\
\bigskip
\textbf{* cmartinezvil@wisc.edu}
\bigskip

\footnotesize{This work has been submitted for publication. Copyright in this work may be transferred without further notice, and this version may no longer be accessible.}

\end{flushleft}

\section*{Abstract}
A theoretical framework is developed for understanding the transient growth and propagation characteristics of thermodynamically coupled, meridional mode-like structures in the tropics. The model consists of a Gill-Matsuno type steady atmosphere under the longwave approximation coupled via a wind-evaporation-sea surface temperature (WES) feedback to a ``slab'' ocean model. When projected onto basis functions for the atmosphere  the system simplifies to a non-normal set of equations that describes the evolution of individual sea surface temperature (SST) modes, with clean separation between symmetric and anti-symmetric modes. The following major findings result from analysis of the system: \emph{(i)} a transient growth process exists whereby specific SST modes propagate toward lower order modes at the expense of the higher-order modes; \emph{(ii)} the same dynamical mechanisms govern the evolution of symmetric and anti-symmetric SST modes except for the lowest-order wave number, where for symmetric structures the atmospheric Kelvin wave plays a critically different role in enhancing decay; and \emph{(iii)} the WES feedback is positive for all modes (with a maximum for the most equatorially confined antisymmetric structure) except for the most equatorially confined symmetric mode where the Kelvin wave generates a negative WES feedback. Taken together, these findings explain why equatorially anti-symmetric ``dipole''-like structures may dominate thermodynamically coupled ocean / atmosphere variability in the tropics. The role of non-normality as well as the role of realistic mean states in meridional mode variability are discussed.

% now start line numbers
%\linenumbers

\section{Introduction}

Distinct patterns of low frequency tropical/subtropical ocean-atmosphere coupled variability in the Pacific and Atlantic occurs mainly via two distinct feedbacks: the so-called Bjerknes feedback [Bjerknes (1969); though the Bjerknes feedback is now understood to involve additional dynamical processes in the ocean (Trenberth et al. 1998)]\nocite{Bjerknes1969, Trenberth1998} and Wind-Evaporation-SST (WES) feedback \cite{XIE1994, Chang1997}. Both of these feedbacks imply a mutual reinforcement between winds and sea surface temperature (SST), but via fundamentally different mechanisms. The Bjerknes feedback relies on ocean dynamical processes in linking wind anomalies to SST tendencies, and typically results in coupled modes that are largely equatorially symmetric such as the El Ni\~no / Southern Oscillation (ENSO). Outside the equatorial zone but within the tropics surface heat fluxes, especially latent and shortwave, play an increasingly important role in SST variability. There, the WES feedback links surface winds to SST tendency through changes in evaporation rates: if positive SST induced wind anomalies are directed opposite to the mean wind direction, the resulting decrease in total wind speed is associated with a reduction in evaporation, and hence a positive SST tendency. The feedback loop is completed in both cases by SST affecting the winds via deep convection \cite{Gill1980} or boundary layer processes \cite{Lindzen1987, Battisti1999b}. 

The WES feedback has mainly been associated with maintaining equatorially antisymmetric coupled modes [meridional modes; Servain et al. (1999); Chiang and Vimont (2004)]\nocite{Servain1999, Chiang2004} by the following mechanism. Consider an anomalous equatorially antisymmetric SST dipole. The atmospheric response to such a dipole includes anomalous winds blowing from the cold hemisphere into the warm hemisphere. Due in part to the Coriolis effect, these anomalous winds relax the mean easterly trades in the warm hemisphere and strengthen them in the cold hemisphere. By wind modulation of evaporation the SST dipole is reinforced. This mechanism has been used to explain the maintenance of meridional modes of ocean-atmosphere covariability such as the Atlantic Meridional Mode \cite{Moura1981, Chang1997}, and the Pacific Meridional Mode \cite{Chiang2004}. Here, we show that interhemispheric anti-symmetry is not necessary for meridional mode evolution and growth.

The mechanisms of growth of meridional modes have been the subject of several theoretical studies. Some of these studies have focused on modal (normal) growth of the least stable eigenvector of a dynamical system that contains the thermodynamic coupling \cite{Noguchi1998, Xie1999, Xie1999a, Wang2010}, while others have investigated the collective transient (non-normal) growth due to positive interference between the individual modes \cite{Vimont2010, Martinez2016}. The propagation characteristics of these kinds of modes has also been the subject of scholarly interest. Liu and Xie (1994) \nocite{Liu1994} and Xie (1999) \nocite{Xie1999} demonstrate that these coupled anomalies propagate westward and equatorward. The equatorward propagation occurs due to wind anomalies that are centered equatorward of SST anomalies, while the westward propagation occurs due to the westward phasing of the Rossby wave atmospheric response to the SST anomalies \cite{Xie1996}. These propagation characteristics may have an important role in linking subtropical SST anomalies to the development of ENSO in the Pacific. As such, a theoretical understanding of the processes that grow and connect subtropical SST anomalies with the equatorial region may shed light on the nature of the observed connection between the Pacific Meridional Mode and ENSO evolution. As of now this connection is not completely understood, and is an area of active research \cite{Vimont2003, Chang2007, Zhang2009, Larson2013, Lin2015, Thomas2016a}.

Although the WES feedback is usually seen as responsible for equatorially antisymmetric variability, a few studies \cite{Noguchi1998, Xie1999, Vimont2010, Wang2010, Martinez2016} have shown that the WES feedback is also able to maintain equatorially symmetric modes. More recently Clement et al. (2011)\nocite{Clement2011}, using an ensemble of General Circulation Models with ``slab'' ocean models (i.e. no Bjerknes feedback included), showed the existence in the equatorial Pacific of ENSO-like ocean-atmosphere modes maintained only by thermodynamic fluxes. It is possible that such a mode exists in nature, but its manifestation is overshadowed by the Bjerknes feedback.

This paper aims to provide basic insight into the similarities and differences between large-scale equatorially symmetric and antisymmetric (meridional mode-like) modes coupled by the WES feedback. Special attention will be given to analyzing mechanisms of propagation and growth for this two set of modes. While other studies have identified similar structures, propagation, and growth characteristics, this study emphasizes the role of transient processes in growth of meridional modes. As such, unlike the earlier referenced studies that focus on growth of a single eigenvector, we show that the interactions between modes (the transient processes) are critical for understanding meridional mode behavior. We note that we refer to these variations as ?meridional modes? irrespective of their equatorial symmetry properties, and whether the structures are best described as a single mode or a transient process. While the name ?meridional mode? is usually used to refer to variations that are equatorially antisymmetric (e.g. a meridional dipole), its use here in a more general sense is justified (as it will be argued later) as growth and propagation of both equatorially symmetric and antisymmetric modes involves the same dynamics through the WES feedback. Further, the propagation in meridional wavenumber (which corresponds to a meridional propagation in physical space) further justifies the use of the term.

The remainder of the paper is organized as follows. Section 2 describes the model and methods used. Section 3 describes and discusses the results. Finally section 4 will present the conclusions.

\section{Model Description}\label{Model}

The model used herein to investigate equatorially symmetric and antisymmetric variability consists of the atmospheric Gill-Matsuno model \cite{Matsuno1966, Gill1980} coupled to a 
thermodynamic ``slab'' ocean. To simplify the calculations and facilitate the interpretation of the results we will adopt the ``long wave approximation'' \cite{Gill1974, Gill1980} in the atmospheric equations, though in Section \ref{Long_wave} we will relax that assumption. For practical purposes the approximation consists of dropping the atmospheric damping in the $v$ equation so the zonal wind is in geostrophic balance. In non-dimensional form, this model may be written:
\begin{align}
\epsilon u-yv&=-\frac{\partial\phi}{\partial x} \nonumber \\
yu&=-\frac{\partial\phi}{\partial y} \nonumber \\
\epsilon\phi+\frac{\partial u}{\partial x}+\frac{\partial v}{\partial y}&=-K_{q}T \nonumber \\
\frac{\partial T}{\partial t}+\epsilon_{T}T&=\alpha u, \label{model}
\end{align}
where $u$ and $v$ denotes low-level zonal and meridional wind anomalies, $\phi$ low level geopotential anomalies, and $T$ SST anomalies. In this model the atmosphere is diagnostic and completely determined by the underlying SST. The SST affects the atmosphere through deep heating, with coupling parameter denoted as $K_{q}$. The atmosphere affects the SSTs through changes in evaporation due to zonal wind anomalies, with parameter $\alpha$ controlling the strength of that coupling. The parameter $\alpha$ can be derived in a variety of ways: \cite{Czaja2002} and \cite{Vimont2009} derive $\alpha$ using a Taylor expansion of the bulk latent heat flux formula around variations in the zonal wind $u$. Notice that in the presence of mean easterlies ($\alpha>0$), a positive $u$ anomaly implies a reduction in wind speed and hence evaporation, and thus induces a positive SST tendency. We take both $K_{q}$ and $\alpha$ parameters to be constant, though the model could be easily expanded to include a spatial dependence of $K_{q}$ or $\alpha$ (this is discussed further in Section~\ref{Concl_sec2}). Both SST and atmospheric parameters are linearly damped by terms $\epsilon_{T}$ and $\epsilon$ respectively. The effective coupling of the system is given by $K_{q}\alpha$ \cite{Martinez2016}. The parameter values are shown in Table 

\begin{table}[H]
\centering
\begin{tabular}{| l | c |}
\hline
Parameters& Value\\
\hline
\ \ $\rho_{o}$ & $10^{3}\: kg\: m^{-3}$\\
\ \ $c_{o}$& $4.2\times 10^{3} J\: kg^{-1}\: K^{-1}$\\
\ \ $H_{o}$& $50\: m$ \\
\ \ $\hat{\alpha}$& $10\frac{J}{m^3}$\\
\ \ $\alpha$& $\frac{\hat{\alpha}}{\rho_{o}c_{o}H_{o}}$\\
\ \ $K_{q}$& $1.7\frac{m^2}{s^3K}$\\
\ \ $\epsilon$& $2\: d^{-1}$ \\
\ \ $\epsilon_{T}$& $120\: d^{-1}$\\
\ \ $\sigma_{0}$& 4.83 (non-dim)\\
\hline
\end{tabular}\label{tab1}
\caption{Model parameters}
\end{table}

\noindent 1. As we are interested in large scale variability, the Gill-Matsuno atmospheric model is appropriate, though Vimont 2010\nocite{Vimont2010} investigates a similar model framework using both the Gill-Matsuno (no long wave approximation used) and Battisti 1999\nocite{Battisti1999b} reduced-gravity boundary layer models. Remaining details of the model are described in Vimont 2010\nocite{Vimont2010} and Martinez-Vilallobos and Vimont 2016\nocite{Martinez2016}.

Note that in (\ref{model}) the SST tendency is coupled to the atmosphere only through zonal wind anomalies $u$, and that the steady atmospheric response can in turn be expressed entirely in terms of temperature. 
We express the atmosphere in terms of the SST alone by decomposing the SST in the meridional direction using parabolic cylinder functions $\psi_{m}(y)$ (see appendix I):
\begin{equation}
T(t,y,x)=\sum_{m=0}^{\infty}T_{m}(t)\psi_{m}(y)\exp{(ikx)}, \label{decomposition}
\end{equation}
where $k$ is the zonal wavenumber considered. A particular parabolic cylinder function $\psi_{m}$ has $m$ zeros, has the symmetry of $m$ ($m$ even corresponds to equatorially symmetric structures, and $m$ odd to equatorially antisymmetric structures), and has loadings increasingly further from the equator as $m$ increases (see figure \ref{Hermite}). So, a low latitude SST signal is dominated by low $m$ modes, while a high latitude signal is increasingly dominated by high $m$ modes. 

The atmosphere may be written in terms of the SSTs (see appendix II) as:
\begin{align}
u(t,y,x)&=\frac{1}{2}(q_{0}(t)\psi_{0}(y)+\sum_{m=1}^{\infty}q_{m+1}(t)(\psi_{m+1}(y)-\sqrt{\frac{m+1}{m}}\psi_{m-1}(y)))exp(ikx)\nonumber\\
v(t,y,x)&=\sum_{m=0}^{\infty}\sqrt{\frac{1}{2(m+1)}}(K_{q}T_{m+1}(t)+(\epsilon+ik)q_{m+1}(t))\psi_{m}(y)exp(ikx)\nonumber\\
\phi(t,y,x)&=\frac{1}{2}(q_{0}(t)\psi_{0}(y)+\sum_{m=1}^{\infty}q_{m+1}(t)(\psi_{m+1}(y)+\sqrt{\frac{m+1}{m}}\psi_{m-1}(y)))exp(ikx) \label{atm}
\end{align}
where $q_{0}$ and $q_{m+1}$ are the amplitudes of the atmospheric Kelvin wave and $m$'th Rossby wave respectively. Here, we use the terms ?Kelvin Wave? and ?Rossby Wave? in the same manner as Gill (1980)\nocite{Gill1980}: to describe features of the steady zonal and meridional atmospheric response to a specific heating structure (in our case, $K_{q}T$). The amplitudes of these waves are determined by the SST as::
\begin{align}
q_{0}&=-K_{q}\frac{T_{0}}{\epsilon+ik}\nonumber\\
q_{m+1}&=K_{q}(\frac{\sqrt{m(m+1)}T_{m-1}+mT_{m+1}}{-(2m+1)\epsilon+ik}) \label{Kelvin_Rossby}
\end{align}

\begin{figure}[H] %Forced here 1
\centerline{\includegraphics[width=6in]{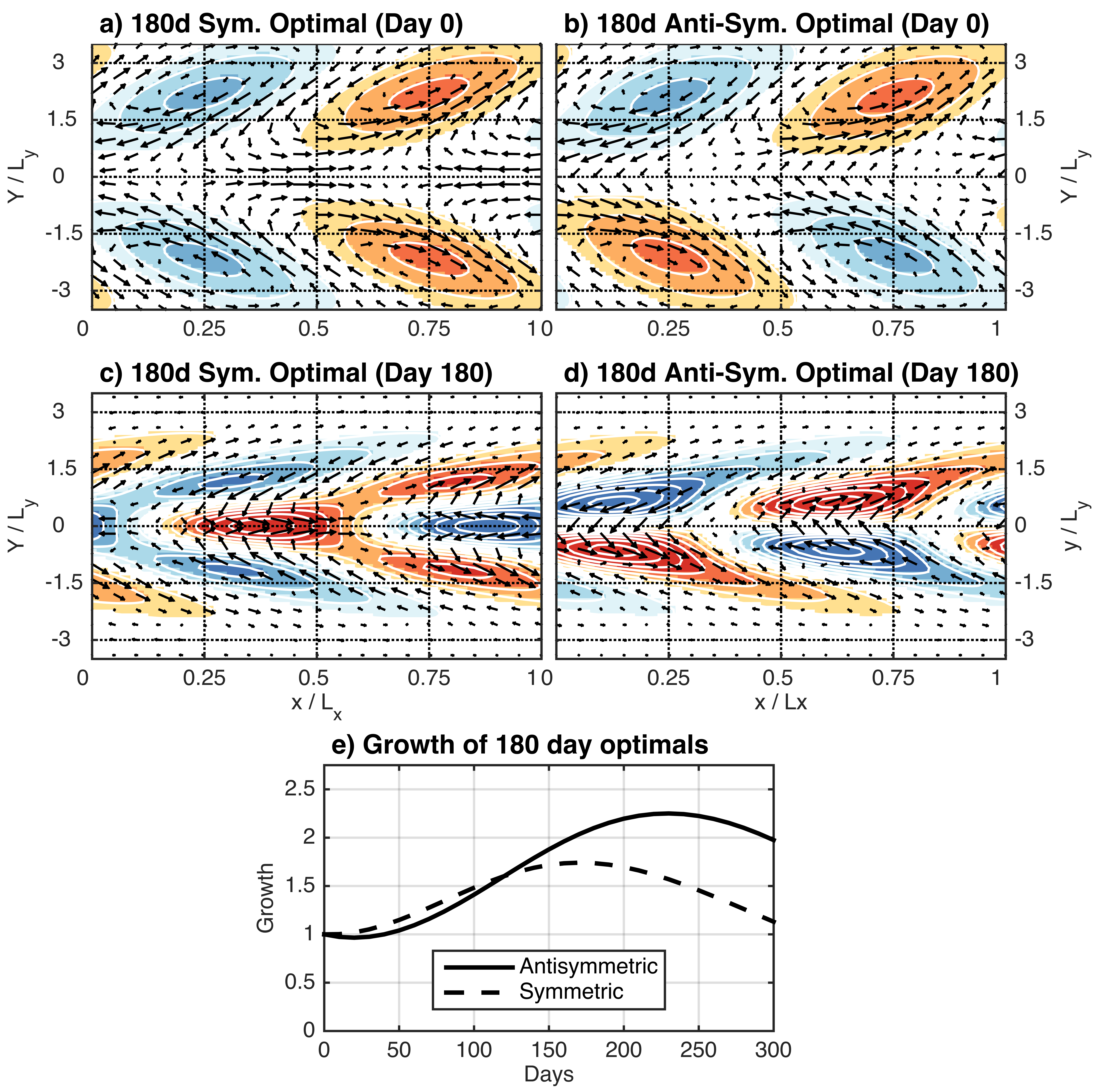}}
\caption{\textbf{(a)} Optimal symmetric and \textbf{(b)} anti-symmetric initial conditions that maximize growth at 180 days. The 180 day evolution of these initial conditions are shown in \textbf{(c)} for the symmetric optimal and \textbf{(d)} for the anti-symmetric optimal. \textbf{(e)} Growth of both optimal initial conditions (note that only SST contributes to state growth). For these plots the wavelength $L_{x}=120^{o}$, and $L_{y}$ is the equatorial radius of deformation ($\sim 10^{o}$ for the parameters in table 1). Units are arbitrary, but consistent for all panels. SST is represented by shading, and low level winds by the arrows.}\label{Sym_anti_180day} %%%From optimal_structures_v_DJV.m
\end{figure}

\noindent Note that the denominator in (\ref{Kelvin_Rossby}) includes a term related to the inverse of the non-dimensional group velocity for non-dispersive (due to the long wave approximation) Kelvin ($+1$) or Rossby ($-(2m+1)$) waves \cite{Gill1980}. Again, as in Gill (1980), the group velocity determines the meridional extent of the response, and in our case the zonal phase difference between the heating (i.e. SST) and the zonal wind component of the atmospheric response (which determines the sign of the WES feedback, and depends on the angle of the complex denominator). These terms in the denominator will prove useful in tracking the role of Rossby or Kelvin waves in the system evolution.

Using (\ref{atm}) and (\ref{Kelvin_Rossby}) in the SST equation for (\ref{model}) we may write an equation for a particular $T_{m}$ just in terms of SST modes.
\begin{align}
\frac{\partial T_{m}}{\partial t}&=\frac{1}{2}K_{q}\alpha[\frac{\sqrt{(m-1)m}}{-(2m-1)\epsilon+ik}T_{m-2}+(\frac{m-1}{-(2m-1)\epsilon+ik}-\frac{m+2}{-(2m+3)\epsilon+ik})T_{m} \nonumber\\&-\frac{\sqrt{(m+1)(m+2)}}{-(2m+3)\epsilon+ik}T_{m+2}]-\epsilon_{T}T_{m}.\label{Tm_eqn}
\end{align}
Also an equation for the squared amplitude of a particular mode can be derived (see appendix III equation \ref{growth_eqn}). We defer the analysis of the terms in these equations to section \ref{Results2}. For now, it is important to notice that the effect of the WES feedback in the evolution of SST anomalies is enclosed in the $K_{q}\alpha$ term, and the SST damping is just the last term. Also, (\ref{Kelvin_Rossby}) and (\ref{Tm_eqn}) shows explicitly that $T_{m}$ is affected by Rossby wave $m-1$ (or Kelvin wave for $T_{0}$) and $m+1$ (note the $2m-1$ or $2m+3$ terms in the denominator, which relates to the Rossby wave group velocity as discussed above), and that symmetric and antisymmetric modes do not interact. This is a consequence of $K_{q}$ and $\alpha$ being symmetric. Vimont 2010\nocite{Vimont2010}, and Martinez-Villalobos and Vimont 2016\nocite{Martinez2016} have studied how meridional variations in the coupling affect thermodynamically coupled variability. 

Because symmetric and antisymmetric modes do not interact, we can analyze them separately. Figure \ref{Sym_anti_180day} shows an example of equatorially symmetric and antisymmetric structures and how they evolve in time. Also plotted is their growth (measured as the basin integrated SST variance, see equation \ref{Tm2}) through the WES feedback. These structures correspond to the leading equatorially symmetric and antisymmetric initial conditions that maximize growth \cite{Farrell1996, Vimont2010} at 180 days of evolution under parameters shown in Table 1 (Appendix IV includes a discussion of growth optimization). The important point for the present discussion is to point out that both symmetric and antisymmetric patterns grow over time (although the antisymmetric pattern grows more), SST structures propagate equatorward and westward for both patterns, and there are processes responsible for the propagation and growth that seemingly work in a similar way for both patterns, namely positive (negative) zonal wind anomalies over positive (negative) SST anomalies. 

\section{Results and Discussion}
\label{Results}

\subsection{Low and high SST modes}
\label{Results1}

In this section we will analyze the evolution of a particular $T_{m}$ mode from (\ref{Tm_eqn}) for low and high $m$ SST modes. We show that a particular mode will propagate ``outward'' to higher and lower order modes, while retaining symmetry. This continues until the lowest-order mode, which excites either the Kelvin wave (symmetric) or mixed Rossby-Gravity wave (anti-symmetric), with fundamental consequences on the growth rate of symmetric or anti-symmetric structures. 

From (\ref{Tm_eqn}) we notice that for $m\geq 2$ both symmetric ($m$ even) and antisymmetric ($m$ odd) SST modes evolve in a qualitatively similar way, via excitation of the $(m+1)$ and $(m-1)$ Rossby waves. Note that the $T_{m+2}$ structure excites the $(m+1)$ and $(m+3)$ Rossby waves (see \ref{Kelvin_Rossby}); the former includes $u$ anomalies that project onto $\psi_{m}$ [last term multiplying $K_q \alpha$ in \ref{Tm_eqn}]. Similarly, the $T_{m-2}$ structure excites the $(m-3)$ and $(m-1)$ Rossby waves (see \ref{Kelvin_Rossby}); the latter includes $u$ anomalies that project onto $\psi_{m}$ [first term multiplying $K_q \alpha$ in \ref{Tm_eqn}]. And finally, the $T_{m}$ structure itself excites both $(m-1)$ and $(m+1)$ Rossby waves, which both include $u$ anomalies that project onto $\psi_{m}$ [middle term multiplying $K_q \alpha$ in \ref{Tm_eqn}]. This demonstrates that for $m\geq 2$ a particular symmetric or antisymmetric mode will propagate ``outward'' (toward the next higher and next lower symmetric or antisymmetric mode, respectively) through Rossby wave excitation. In physical space, the propagation toward different modes represents a meridional propagation toward a higher latitude variation (for $m$ increasing) or toward more equatorially confined variability (for $m$ decreasing).

For $m=0$ and $m=1$ the situation is different. In this case the evolution of both $T_{0}$ and $T_{1}$ modes are dictated by qualitatively different equations,
\begin{align}
\frac{\partial T_{0}}{\partial t}&=\frac{1}{2}K_{q}\alpha[(\frac{-1}{\epsilon+ik}-\frac{2}{-3\epsilon+ik})T_{0}-\frac{\sqrt{2}}{-3\epsilon+ik}T_{2}]-\epsilon T_{0}\nonumber\\
\frac{\partial T_{1}}{\partial t}&=\frac{1}{2}K_{q}\alpha[(\frac{-3}{-5\epsilon+ik}T_{1}-\frac{\sqrt{6}}{-5\epsilon+ik}T_{3}]-\epsilon_{T}T_{1}. \label{T0_T1}
\end{align}
The evolution of the $T_{0}$ term is affected by both atmospheric Kelvin wave and the first symmetric Rossby wave, whereas the $T_{1}$ term is affected by just the first antisymmetric Rossby wave [the next-lower anti-symmetric mode is the mixed Rossby-Gravity wave, for which $u$ anomalies under the long-wave approximation are strictly zero as shown in (\ref{Kelvin_Rossby}) (i.e. $q_{1}=0$)].

Taken together, the full evolution from $m\geq 2$ toward $m=0,1$ modes illustrates two important characteristics of thermodynamically coupled variability. First, for $m\geq 2$ the dynamics of both equatorially symmetric and antisymmetric SST patterns is essentially the same. Second, the dynamics of the system only differ for the lowest order modes $m=0,1$, where the Kelvin (for $m=0$) and mixed Rossby-Gravity (for $m=1$) modes have fundamentally different effects. Note from Fig.~\ref{Sym_anti_180day} that the total growth for the symmetric and antisymmetric patterns is quite different. We will demonstrate below that the reason for this difference lies in the atmospheric Kelvin wave. 

\subsection{Analysis of terms in SST equation}
\label{Results2}

In order to analyze (\ref{Tm_eqn}) in the parameter space considered herein we rewrite it in terms of the ratio between the wavenumber and atmospheric damping ($\nu$), and the ratio between the coupling and damping terms ($\sigma$ the stability parameter):
\begin{align}
\nu&=\frac{k}{\epsilon} \nonumber\\
\sigma&=\frac{K_{q}\alpha}{\epsilon_{T}\epsilon},
\end{align}
as
\begin{equation}
\frac{\partial T_{m}}{\partial t}=\frac{1}{2}\epsilon_{T}(-h(m-1,\nu)\sigma T_{m-2}+f(m,\sigma,\nu)T_{m}+h(m+1,\nu)\sigma T_{m+2}), \label{h_f_Tm}
\end{equation}
where the functions introduced are defined as:
\begin{align}
h(m,\nu)&=\frac{\sqrt{m(m+1)}}{(2m+1)-i\nu}\nonumber \\
f(m,\sigma,\nu)&=g(m,\nu)\sigma-2\nonumber \\
g(m,\nu)&=\frac{((2m+1)-3i\nu)}{(2m-1)(2m+3)-2i\nu(2m+1)-\nu^2}. \label{h_f_functions}
\end{align}
The function $h(m,\nu)$, the exchange function, encodes part of the influence of Rossby wave $m$ [the atmospheric Kelvin wave does not participate in the exchange as $h(-1,\nu)=0$] via connecting $T_{m-2}$ and $T_{m+2}$ to $T_{m}$, and its real part is always positive. Notice that the sign of $h(m,\nu)$ in (\ref{h_f_Tm}) shows that both $T_{m-2}$ and $T_{m+2}$ terms affect $T_{m}$ 

\begin{figure}[H]
\centerline{\includegraphics[width=6in]{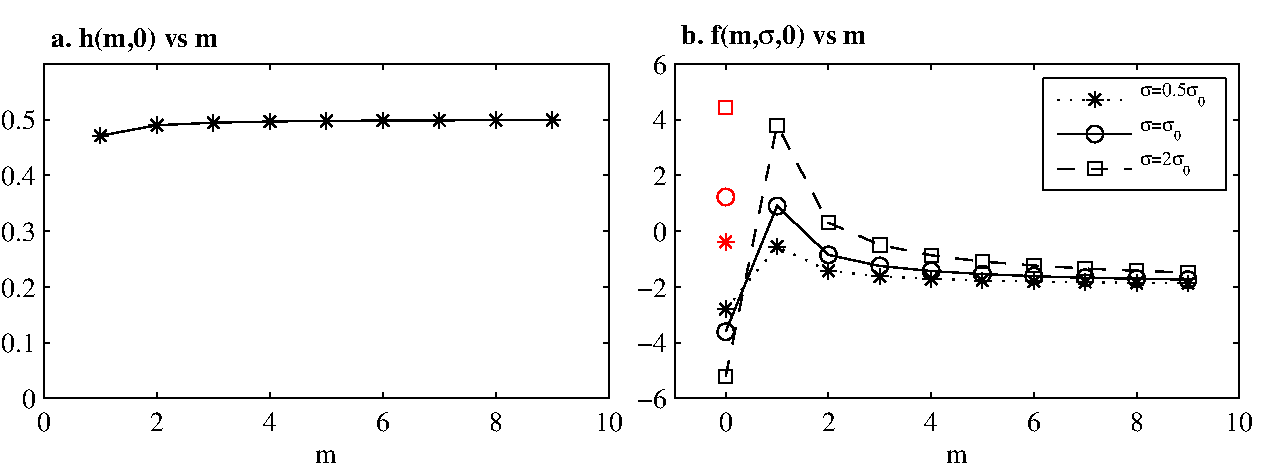}}
\caption{\textbf{(a)} The exchange function $h(m,0)$ as a function of SST mode $m$. This function monotonically increases towards its asymptotic value of $\frac{1}{2}$. \textbf{(b)} The growth function $f(m,\sigma,0)$ as a function of $m$ for different values of the stability parameter $\sigma$. Values in red show $f(m=0,\sigma,0)$ when the atmospheric Kelvin wave is suppressed.} \label{h_and_f}   %%%From k_zero_case.m
\end{figure}

\noindent evolution in different directions (opposite polarity). This will prove important in explaining the propagation characteristics and transient growth of the modes. 

Function $f(m, \sigma, \nu)$ contains the effect of $T_{m}$ on itself. Here, the function $g(m,\nu)\sigma$ contains part of the WES feedback influence on growth (if $g>0$ it contributes to $T_{m}$ growth, and if $g<0$ it contributes to $T_{m}$ decay) and the $-2$ term corresponds to the SST damping. Notice that the instantaneous growth or decay of a single $T_{m}$ mode depends on the balance between the WES feedback [$g(m,\nu)\sigma$] and the SST damping ($-2$), where here we distinguish the WES feedback as including \emph{only} the wind-induced change in surface latent heat flux.

\subsection{Long zonal wavelength limit}
\label{Results3}

As a starting point we will study the long zonal wavelength ($k\rightarrow 0$) limit, i.e. $\nu=0$. This limit will prove useful in understanding the more general variability for long but finite zonal wavelengths. In this case (\ref{Tm_eqn}) (also \ref{h_f_Tm}) is reduced to
\begin{equation}
\frac{\partial T_{m}}{\partial t}=\frac{1}{2}\epsilon_{T}(-h(m-1,0)\sigma T_{m-2}+f(m,\sigma,0)T_{m}+h(m+1,0)\sigma T_{m+2}) \label{k0_eqn}.
\end{equation}
Functions $h$ and $f$ in this limit are purely real, as in this limit atmospheric waves are zonally in phase with the SST:
\begin{align}
h(m,0)&=\frac{\sqrt{m(m+1)}}{(2m+1)}\nonumber \\
f(m,\sigma,0)&=g(m,0)\sigma-2=\frac{(2m+1)\sigma}{(2m-1)(2m+3)}-2. \label{f_h_k0}
\end{align}
In this case, the equation for a particular mode growth (see appendix III equation \ref{growth_eqn}) is vastly simplified to
\begin{equation}
\frac{\partial |T_{m}|^2}{\partial t}=\epsilon_{T}(-h(m-1,0)\sigma T_{m-2}T_{m}+f(m,\sigma,0)|T_{m}|^2+h(m+1,0)\sigma T_{m+2}T_{m}).\label{Tm_growth_k0}
\end{equation}
Figure \ref{h_and_f}a shows $h(m,0)$ as a function of mode $m$. This shows that $h$ is a very flat function of $m$ for $m>1$ ($h(m-1,0)=0$ for $m=0,1$). The difference in value between $h(m+1,0)$ and $h(m-1,0)$ is positive and small, and gets progressively smaller as $m$ increases ($\frac{dh(m,0)}{dm}\sim\frac{1}{16m^2}>0$). This implies that if $T_{m-2}$ has the same magnitude as $T_{m+2}$ their effect on $T_{m}$ growth roughly cancels out, though the effect of $T_{m+2}$ slightly dominates. This dominant influence of the mode ($m+2$) on the lower-order $m$ helps explain the equatorward propagation (i.e. propagation towards lower $m$) seen in figure \ref{Sym_anti_180day}.

Next, consider the growth equations for modes $m-2$ and $m+2$
\begin{align}
\frac{\partial |T_{m-2}|^2}{\partial t}&=\epsilon_{T}(-h(m-3,0)\sigma T_{m-4}T_{m}+f(m-2,\sigma,0)|T_{m-2}|^2+h(m-1,0)\sigma T_{m}T_{m-2}).\label{Tm_2_growth_k0}\\
\frac{\partial |T_{m+2}|^2}{\partial t}&=\epsilon_{T}(-h(m+1,0)\sigma T_{m}T_{m+2}+f(m+2,\sigma,0)|T_{m+2}|^2+h(m+3,0)\sigma T_{m+4}T_{m+2}).\label{Tm+2_growth_k0}.
\end{align}
We notice that the third term in equation \ref{Tm_2_growth_k0} is exactly equal to the first term in equation \ref{Tm_growth_k0}, but with opposite sign; similarly with the first term of equation \ref{Tm+2_growth_k0} and third term in equation \ref{Tm_growth_k0}. Without loss of generality also consider that $T_{m-2}$, $T_{m}$ and $T_{m+2}$ all start positive. This shows that a positive $T_{m+2}$ will grow $T_{m}$ amplitude, while at the same time $T_{m}$ will decrease $T_{m+2}$ amplitude. Then $T_{m}$ will increase $T_{m-2}$ amplitude while $T_{m-2}$ will damp $T_{m}$. As a consequence of this process $T_{m+2}$ grows $T_{m}$ that then grows $T_{m-2}$. This, at the core, is the mechanism that explains equatorward propagation (propagation towards lower order modes) of thermodynamically coupled variability. 

The $f(m,\sigma,0)$ function in (\ref{k0_eqn}) and (\ref{f_h_k0}) is key in the prospect of long term growth of a particular mode. This function depends on the particular mode $m$ and the ratio between coupling and damping, the stability parameter $\sigma$. A particular mode $m$ will grow (at least for some time) if $f(m,\sigma,0)>0$, otherwise it decays. Figure 2b shows how $f(m,\sigma,0)$ varies with $m$ (for the first 10 $m$ modes), for different values of the stability parameter. Here $\sigma_{0}=4.83$ using the standard values shown in table 1. For the $m$ values of interest $g(m,0)$ has a pole at $m=\frac{1}{2}$ implying, as shown in figure \ref{h_and_f} for 3 different values of $\sigma$, a very different behavior for modes $m=0$, and $m=1$. In this limit $g(m,0)$ peaks at $m=1$, is positive for $m\geq 1$, and is negative just when $m=0$. This shows that the WES feedback is positive (it works towards amplifying an initial $T_{m}$ anomaly) for all $m$ modes, except $m=0$. The function $g(m,0)$ gets monotonically smaller as $m$ increases, implying that the WES feedback gets progressively less effective as we move to higher latitudes, consistent with \cite{Vimont2010}. 

For $m\geq 1$ there is a tension in function $f(m,\sigma,0)$ between a positive WES feedback represented by the $g(m,0)\sigma$ term, that acts to enhance $|T_{m}|^2$ growth, and the SST damping represented by the $-2$ term, that acts to damp SST. When $g(m,0)\sigma$ is greater than 2 there is a prospect for transient growth. In the cases shown in figure 2b, $f(m,\frac{1}{2}\sigma_{0},0)<0$ for all $m'$s so the system simply decays, though at a rate slower than $\epsilon_{T}^{-1}$. For $\sigma=\sigma_{0}$, $f(m,\sigma_{0},0)>0$ just for $m=1$ so the system will grow, at least for  some time, just when the SSTs project onto $\psi_{1}(y)$. Finally, for $\sigma=2\sigma_{0}$ $f(m,2\sigma_{0},0)>0$ for $m=1,2$, so the SSTs will grow if they project onto $\psi_{1}$ and $\psi_{2}$, but still the greatest growth will be achieved when they project onto $\psi_{1}$. For $m=0$ $g(m,0)\sigma<0$, so the WES feedback turns into a negative feedback, and the system is additionally damped for that mode.

The reason why $T_{0}$ is damped by, and $T_{1}$ grows through the WES feedback lies on the differences between atmospheric Kelvin and Rossby waves. Consider an equatorially symmetric zonally homogeneous positive SST anomaly confined close to the equator. The atmospheric response includes (see equation \ref{Kelvin_Rossby} for $q_{0}$ and $q_{2}$) a positive zonal wind anomaly $u_{R}$ associated with the first symmetric Rossby wave on top of the anomaly (recall, there is zero zonal phase difference between SST and the atmosphere in the 

\begin{figure}[H]
\centerline{\includegraphics[width=3.125in]{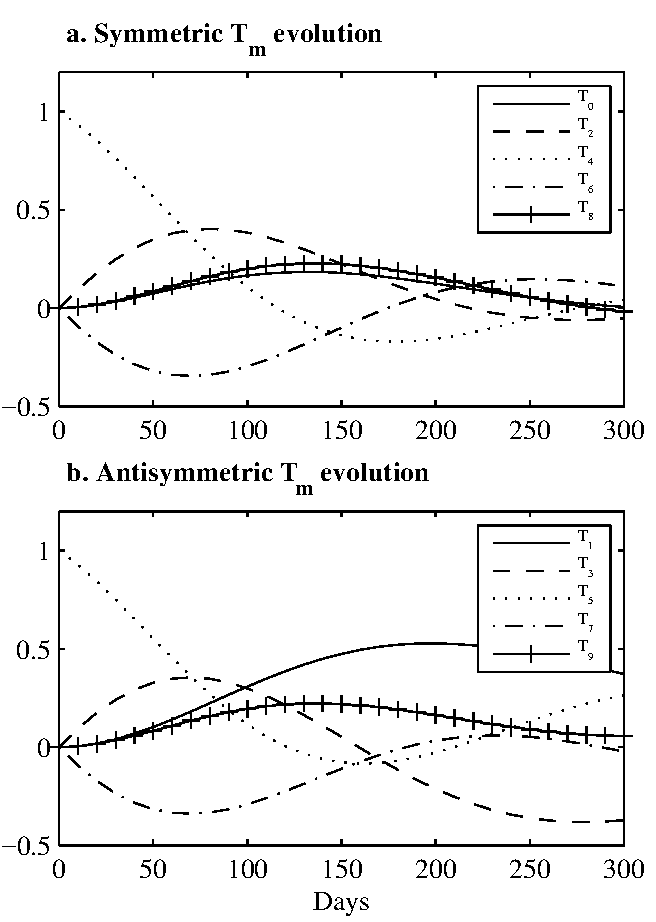}}
\caption{\textbf{(a)} Evolution of individual $T_{m}$ coefficients for the symmetric initial condition $T(0)=\psi_{4}(y)$. \textbf{(b)} Evolution of individual $T_{m}$ coefficients for the antisymmetric initial condition $T(0)=\psi_{5}(y)$. Both panels are shown for stability parameter $\sigma_{0}$ and for $k = 0$.} \label{Tn_evolution} 
\end{figure}

\noindent $\nu=0$ case), and a negative zonal wind anomaly $u_{K}$ associated with the Kelvin wave. Since the mean zonal winds are negative along the equator ($\alpha>0$), the Rossby response acts to decrease evaporation further increasing the positive SST anomaly, whereas the Kelvin wave acts to increase evaporation, damping the SST anomaly. 

This result implies a very different situation between symmetric and antisymmetric SST patterns with respect to growth. A subtropical \emph{antisymmetric} pattern will propagate equatorward (towards lower $m$) when it will finally project onto $\psi_{1}$. At that point the system will grow while at the same time causing a $T_{3}$ tendency in the opposite direction (see equation \ref{Tm_growth_k0} for $m=1$). As $T_{3}$ grows in the opposite direction, it ultimately counteracts the growth of $T_{1}$ causing the entire system to decay back to zero. On the other hand, a subtropical equatorially \emph{symmetric} pattern will also propagate towards lower $m$, finally projecting onto $\psi_{0}$ where it will be additionally damped by the WES feedback due to the atmospheric Kelvin wave effect (see equation \ref{T0_T1} with $k=0$). 

To illustrate how the Kelvin wave affects growth, we will consider the $\sigma=\sigma_{0}$ case from here on. In this case just the $m=1$ mode has prospective growth. For the rest of the paper we will use the first 10 $T_{m}$ modes in our calculations (5 symmetric and 5 antisymmetric). We have tested that results do not change qualitatively by increasing the number of modes retained. Figure 3 shows how this asymmetry in the evolution of equatorially symmetric and antisymmetric SST anomalies in this limit plays out. To make things easier to interpret, the system will be initialized as $T(0,y)=\psi_{4}(y)$ in the symmetric case, and $T(0,y)=\psi_{5}(y)$ for the antisymmetric case. In accordance with (\ref{k0_eqn}) both patterns start very similarly. The symmetric (anti-symmetric) pattern 

\begin{figure}[H]
\centerline{\includegraphics[width=3.125in]{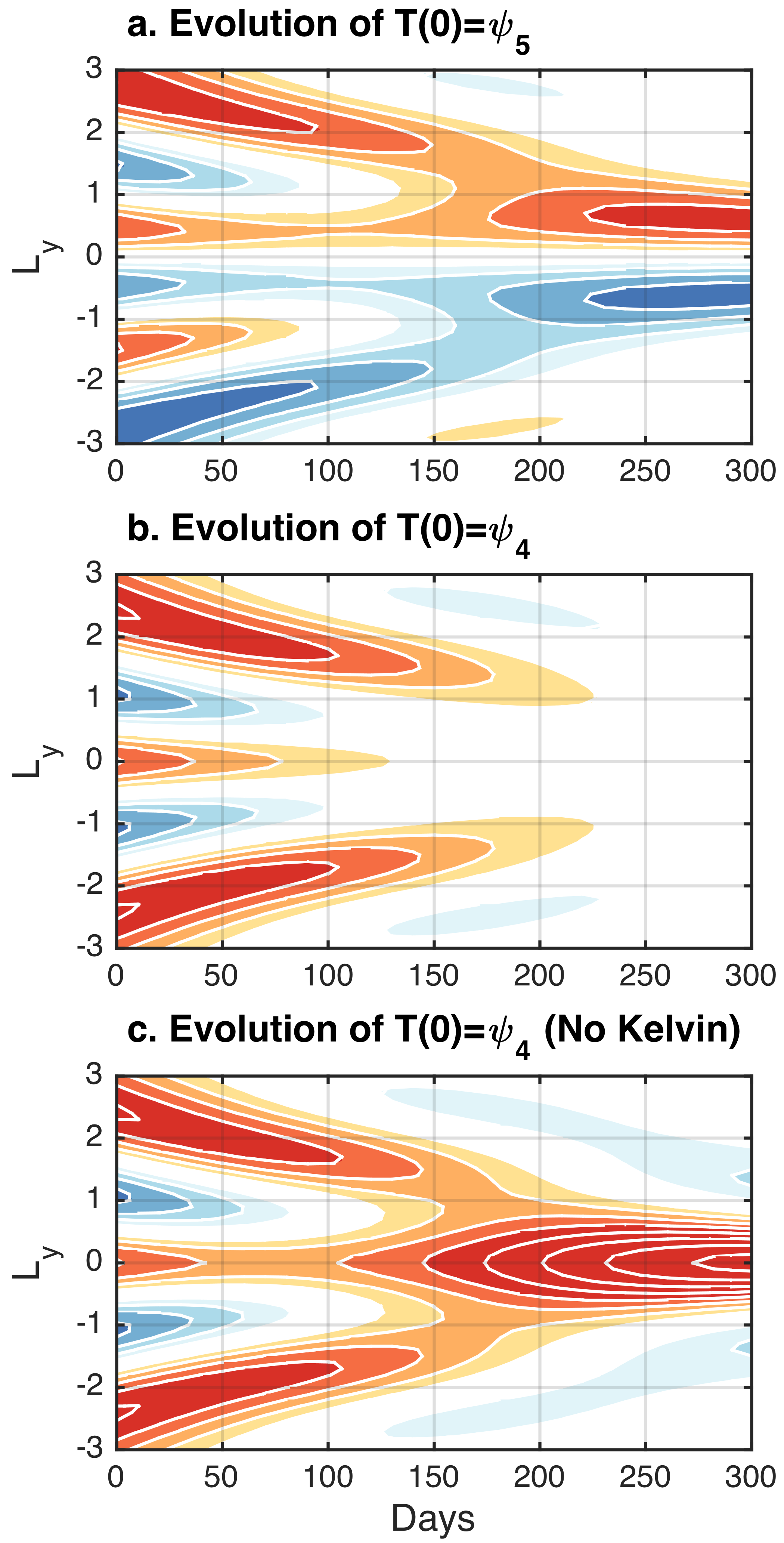}}
\caption{Latitude-time evolution of SST from 0 to 300 days.  \textbf{(a)} Evolution of the antisymmetric initial condition $T(0)=\psi_{5}(y)$. \textbf{(b)} Evolution of the symmetric initial condition $T(0)=\psi_{4}(y)$. \textbf{(c)} Same as \textbf{(b)} except that the atmospheric Kelvin wave is suppressed.  Shading represents SST. In this $k=0$ case the evolution does not depend on longitude.} \label{Hovmoller}
\end{figure}

\noindent generates a positive $T_{2}$ ($T_{3}$) response, and a similar but albeit smaller, negative $T_{6}$ ($T_{7}$) response, while being damped by these modes in the process. Both positive signals continue propagating equatorward, losing energy in the process until the antisymmetric SST pattern projects onto $\psi_{1}$, where the system grows for some time, and the symmetric SST pattern project onto $\psi_{0}$, where the system decays even more rapidly. Notice how similar figures \ref{Tn_evolution}a and \ref{Tn_evolution}b are, except in the $T_{0}$ and $T_{1}$ evolution.

This general pattern is confirmed in figure \ref{Hovmoller}. This figure shows how the symmetric initial structure $T(0,y)=\psi_{4}(y)$ and antisymmetric initial structure $T(0,y)=\psi_{5}(y)$ evolve spatially in the meridional direction from 0 to 300 days. Also shown in this figure 
\begin{figure}[H]
\centerline{\includegraphics[width=3.125in]{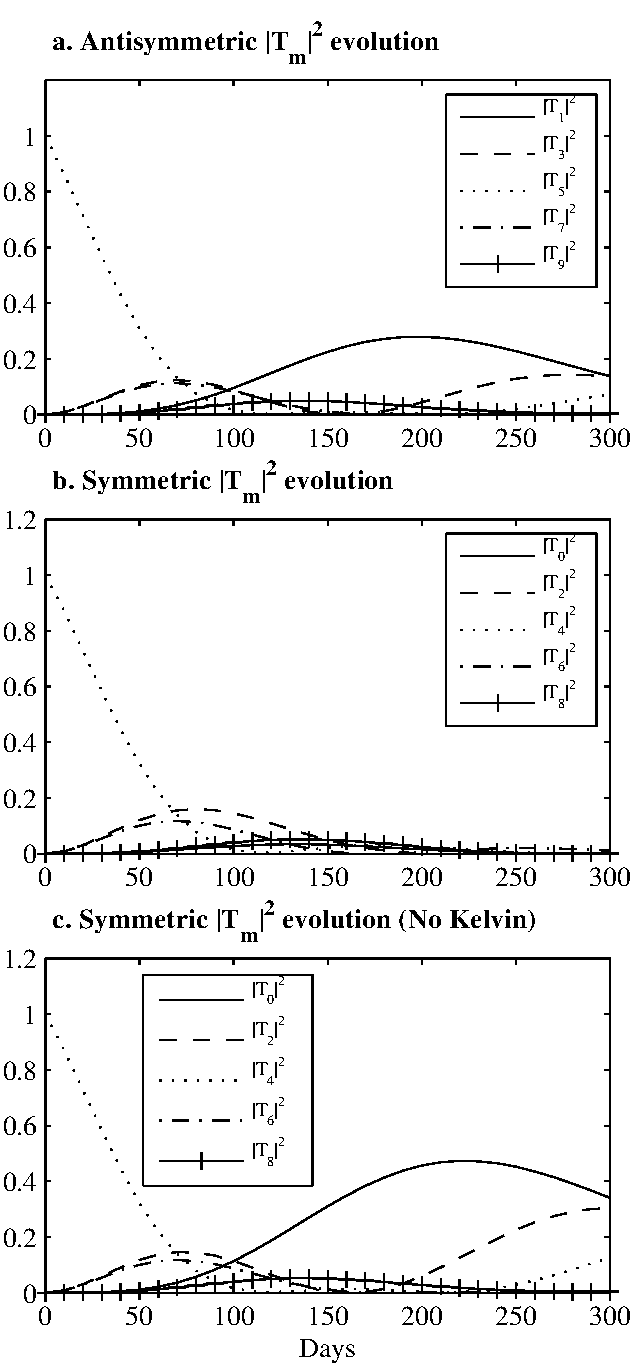}}
\caption{$|T_{m}|^2$ evolution for: \textbf{(a)} the antisymmetric initial condition $T(0)=\psi_{5}(y)$, \textbf{(b)} the symmetric initial condition $T(0)=\psi_{4}(y)$, and \textbf{(c)} $T(0)=\psi_{4}(y)$ [as in \textbf{(b)}] except that the atmospheric Kelvin wave is suppressed. All panels are shown for stability parameter $\sigma_{0}$ and for $k = 0$.} \label{Tn2_evolution} 
\end{figure}
\noindent is how the symmetric pattern would evolve if we artificially suppress the atmospheric Kelvin wave in the SST calculation. This suppression consists of neglecting the $-\frac{1}{\epsilon+ik}$ ($k=0$) term in the $T_{0}$ evolution calculation (see equation \ref{T0_T1}). (This has the effect of changing the functional form of $f$ in equation \ref{f_h_k0} for the no Kelvin wave case --see red dots figure \ref{h_and_f}b--). Notice how all pattern amplitudes look similar at the beginning. They all evolve equatorward, where the major changes are more apparent. The antisymmetric structure decreases in magnitude as it propagates toward the equator from day 0 to about day 150 (notice the decrease in the shading) until the SSTs projects onto $\psi_{1}(y)$ close to the equator where growth increases again (notice the shading starting at day 150). In contrast, Fig.~\ref{Hovmoller}b shows the amplitude of the symmetric pattern simply decaying with time. We compare this pattern with Fig.~\ref{Hovmoller}c which shows how the symmetric structure evolves without the atmospheric Kelvin wave. The damping effect of the Kelvin wave is apparent from about day 100: while the symmetric structure equatorial amplitude continues to decay in Fig.~\ref{Hovmoller}b, the pattern grows at the equator when the Kelvin wave is suppressed (Fig.~\ref{Hovmoller}c). This is a contrast with the similarity between both patterns at the beginning, when the signal had not yet reached the equator. 

Figure~\ref{Tn2_evolution} shows the evolution of the SST modal amplitude $T_{m}^2$ for the same initial conditions considered above, including the evolution of the symmetric initial condition when we suppress the atmospheric Kelvin wave. Not only is the symmetric initial condition able to grow transiently in this case, its growth is actually bigger than the antisymmetric case. This is consistent with the values of the growth function $f(m=0,\sigma_{0},0)$ when we zero out the Kelvin wave (red dots in Fig.~\ref{h_and_f}b). This shows how powerful and determinant the atmospheric Kelvin wave is in affecting thermodynamically coupled mode dynamics. Without its influence thermodynamically coupled symmetric modes would actually grow more than equatorially antisymmetric ones. 

\subsubsection{Total SST growth}
\label{Results3.1}

From the SST equation (equation \ref{model}) we calculate an equation for the growth of total SST anomalies. In this $\nu=0$ limit (see appendix III for the general growth equation), and considering equation \ref{Tm_growth_k0}, this is equal to
\begin{equation}
\frac{\partial <T^2>}{\partial t}=L_{x}\sum_{m=0}^{\infty}\frac{\partial |T_{m}|^2}{\partial t}=L_{x}\epsilon_{T}\sum_{m=0}^{\infty}f(m,\sigma,0)|T_{m}|^2, \label{Total_growth}
\end{equation}
where the symbol $<>$ means integration over the basin considered, and $L_{x}$ is its zonal extension. The exchange terms in equation \ref{Tm_growth_k0} ($\propto h$) get canceled out when we sum all the $\frac{\partial T_{m}^2}{\partial t}$ terms. The total SST growth evolution is described by this equation, together with equation \ref{Tm_growth_k0} that determines the growth of an individual mode. Equations (\ref{Total_growth} and \ref{Tm_growth_k0}) indicate that in the $\nu=0$ limit the role of the exchange function $h(m,0)$ is to propagate the signal towards lower modes (curbing growth in the process, as explained in next section), and the growth function $f(m,\sigma,0)$ determines whether that signal transiently grows or decays.

\begin{figure}[H]
\centerline{\includegraphics[width=3.125in]{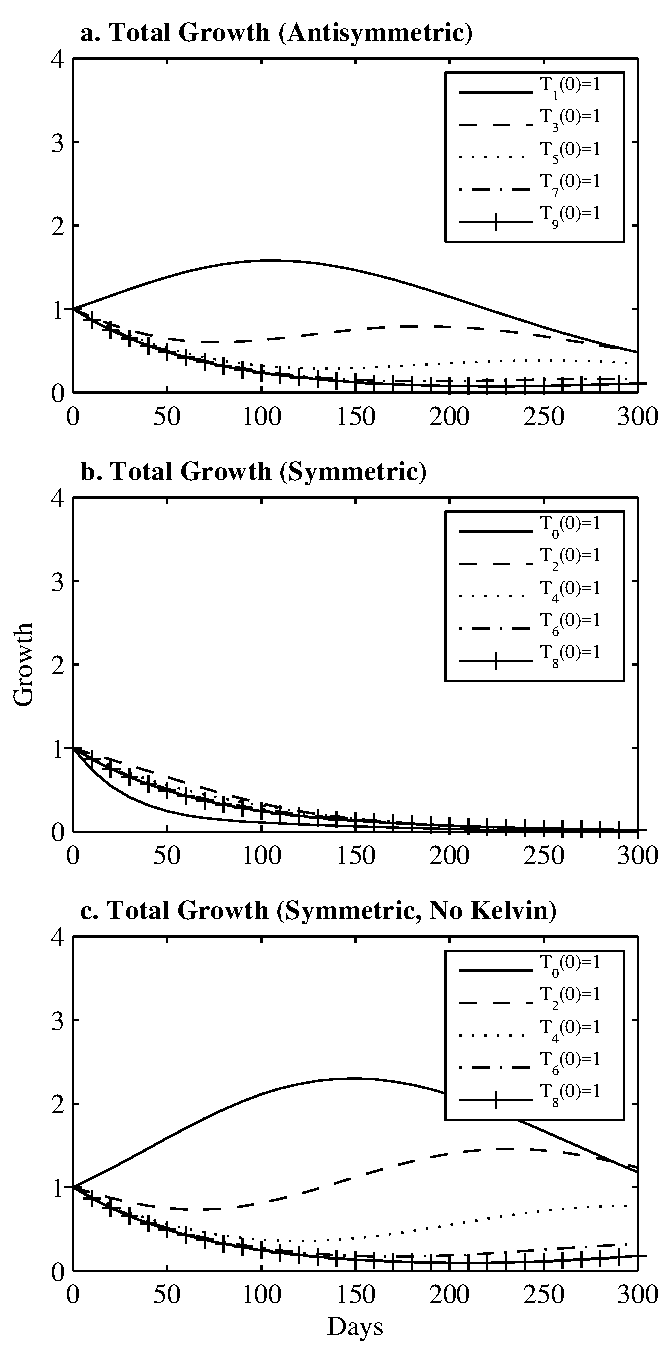}}
\caption{Total SST growth for \textbf{(a)} antisymmetric initial conditions, \textbf{(b)} symmetric initial conditions, and \textbf{(c)} symmetric initial conditions [as in \textbf{(b)}] except that the atmospheric Kelvin wave is suppressed. For panel \textbf{(a)} the lines show the total growth for initial conditions starting at $T_{1}$ (solid), $T_{3}$ (dashed), $T_{5}$ (dotted), $T_{7}$ (dash-dot), and $T_{9}$ ($+$ sign). For panels \textbf{(b)} and \textbf{(c)} the lines show the total growth for initial conditions starting at $T_{0}$ (solid), $T_{2}$ (dashed), $T_{4}$ (dotted), $T_{6}$ (dash-dot), and $T_{8}$ ($+$ sign). All panels are shown for stability parameter $\sigma_{0}$ and for $k = 0$.} \label{Total_no_kelvin}
\end{figure}

Figure \ref{Total_no_kelvin} shows the total growth $<T^2>$ for antisymmetric initial conditions $T(0)=\psi_{1}(y)$, $\psi_{3}(y)$, $\psi_{5}(y)$, $\psi_{7}(y)$, $\psi_{9}(y)$ (fig \ref{Total_no_kelvin}a), symmetric initial conditions $T(0)=\psi_{0}(y)$, $\psi_{2}(y)$, $\psi_{4}(y)$, $\psi_{6}(y)$, $\psi_{8}(y)$ (fig \ref{Total_no_kelvin}b), and the same symmetric initial conditions with the atmospheric Kelvin wave suppressed (fig \ref{Total_no_kelvin}c). We notice that the $T_{1}$ antisymmetric initial condition grows for some time (about $100d$) before it is damped by $T_{3}$ (see \ref{Tm_growth_k0}). All the other antisymmetric initial conditions decay until the signal propagates close enough to the equator that it projects significantly onto $\psi_{1}(y)$, at which point the system experiences some growth, leading to secondary peaks that are less energetic and peak later in time (e.g. around $190d$ for $T_{3}$ and $250d$ for $T_{5}$). The symmetric initial conditions all decay, with $T_{0}$ decaying the fastest (due to the atmospheric Kelvin wave damping effect) and $T_{2}$ the slowest. This is consistent with the values of $f(m,\sigma_{0},0)$ shown in figure \ref{h_and_f}b. When the Kelvin wave is suppressed the system is able to grow more effectively for initial conditions starting near the equator (lower order modes). Without the Kelvin wave, the SSTs are able to grow due to the positive effect of the atmospheric Rossby waves. This is consistent with the value of the growth function $f(m,\sigma_{0},0)$ with the Kelvin wave suppressed (figure \ref{h_and_f}b red dots).

\subsubsection{Normal vs non-normal growth}
\label{Results3.2}

In the limit $\nu\rightarrow 0$ we may write equation \ref{k0_eqn} in matricial form 
\begin{equation}
\frac{\partial\mathbf{T}}{\partial t}=\mathbf{M}\mathbf{T}=-i\omega\mathbf{T}, \label{solution}
\end{equation}
\begin{figure}[H]
\centerline{\includegraphics[width=3.125in]{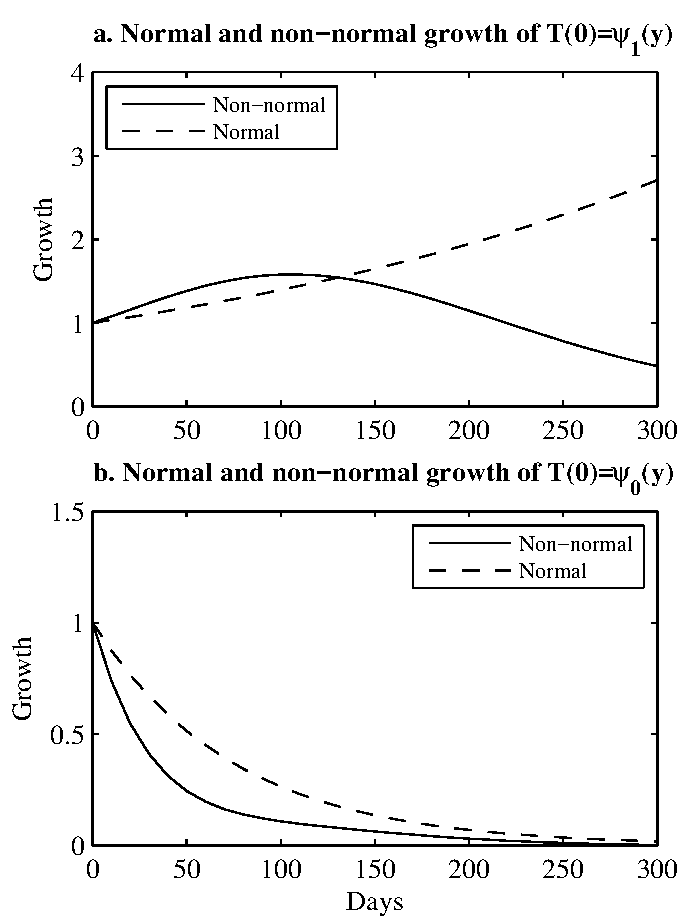}}
\caption{Total SST growth for the non-normal system (solid line; equation~\ref{k0_eqn}), and for the ``normalized'' system (dashed line; equation~\ref{k0_eqn_normal}; see text for details). \textbf{(a)} Growth from antisymmetric $T(0)=\psi_{1}(y)$ initial conditions. \textbf{(b)} Growth from symmetric $T(0)=\psi_{0}(y)$ initial conditions. Both cases are shown for stability parameter $\sigma_{0}$ and for $k = 0$.} \label{non_normal}
\end{figure}
\noindent where $\mathbf{T}=(T_{0},\: T_{1},\: T_{2},\:...)exp(-i\omega t)$, and $\mathbf{M}$ entries are given by
\begin{align}
M_{m,m-2}&=-\frac{1}{2}\epsilon_{T}h(m-1,0)\sigma\nonumber\\
M_{m,m}&=\frac{1}{2}f(m,\sigma,0)\nonumber\\
M_{m,m+2}&=\frac{1}{2}\epsilon_{T}h(m+1,0)\sigma. \label{Mk0}
\end{align}
As previously stated we truncate at $m=m_{T}=9$ in our calculations, so \textbf{M} is a 10 by 10 matrix. The term $M_{9,11}$ in (\ref{Mk0}) is neglected. It has been corroborated numerically (not shown, but similar as explained in section 3e) that this choice does not affect the results. For $\sigma=\sigma_{0}$ the spectrum of $\mathbf{M}$ is such that the system is linearly stable (i.e. $Im(\omega)<0$ for all $\omega$) and non-normal (i.e. $\mathbf{M}^{T}\mathbf{M}\neq\mathbf{M}\mathbf{M}^{T}$), so growth may be achieved transiently \cite{Farrell1996}. This also implies that  $\lim_{t\rightarrow\infty}T=0$ as is the case in previous plots. 

The $\nu\rightarrow 0$ limit provides a convenient way to show the differences between normal and non-normal growth. The system is non-normal due to the exchange function $h$ being non-zero and of opposite sign for $M_{m+2,m}$ and $M_{m,m+2}$ (see \ref{Mk0}). Artificially suppressing non-normality by zeroing out the exchange functions, the equation that determines the evolution of a particular mode amplitude is now equal to (compare to equation \ref{k0_eqn}):
\begin{equation}
\frac{\partial T_{m}}{\partial t}=\frac{1}{2}\epsilon_{T}f(m,\sigma,0)T_{m}. \label{k0_eqn_normal}
\end{equation}
\begin{figure}[H]
\centerline{\includegraphics[width=3.125in]{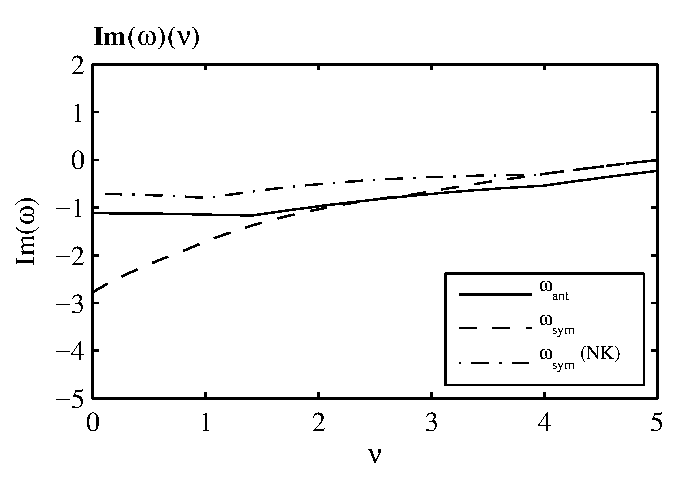}}
\caption{Growth rate ($\times 10^{-3}\frac{s^{-1}}{\sqrt{c\beta}}$) of the least stable eigenmode of $\mathbf{M}$ (equation~\ref{M}) as a function of $\nu$. $Im(\omega)>0$ implies a linearly unstable mode. The parameter $\nu$ is defined as $\nu=\frac{k}{\epsilon}$. For the parameters shown in table 1, a wavelength $L_{x}$ of $120^{o}$ corresponds to $\nu=2.44$. } \label{Im(w)}
\end{figure}
\noindent The equation for total SST growth (equation \ref{Total_growth}) remains unchanged, and actually contains the same information as (\ref{k0_eqn_normal}). Note, however, that in the normal system (no exchange terms) each mode can be integrated individually using (\ref{k0_eqn_normal}) and the result summed to obtain total growth (the integration of equation \ref{Total_growth}). The solution to equation \ref{k0_eqn_normal} is
\begin{equation}
T_{m}(t)=T_{m}(0)\exp{(\frac{1}{2}\epsilon_{T}f(m,\sigma,0)t)}.
\end{equation}
We observe that without the exchange terms the system will just decay or grow indefinitely depending on the sign of $f$. That is, for $\sigma = \sigma_0$ the system will grow indefinitely if the initial SSTs project onto $\psi_{1}(y)$ otherwise it will decay. In reality, the WES feedback coupling is ``non-normal'' so the growth is curbed by the exchange terms. In that case the SSTs will grow for some time if they project onto $\psi_{1}$, but the growth will be limited by the effect of the exchange terms.

Figure \ref{non_normal} shows the total growth for both normal (\ref{k0_eqn_normal}) and non-normal (\ref{k0_eqn}) cases for the antisymmetric initial condition $T(0)=\psi_{1}(y)$ and symmetric initial condition $T(0)=\psi_{0}(y)$. In the normal case the antisymmetric initial condition grows exponentially, and the symmetric initial condition decays exponentially. In the non-normal case the growth of the antisymmetric initial condition initially exceeds growth of the normal system due to growth of $T_3$ and higher modes, but is eventually curbed by the interaction with $T_{3}$ (the exchange term $\propto h(2,0)\sigma_{0}T_{1}T_{3}$ in equation \ref{k0_eqn}). Similar arguments hold for the symmetric case $T_{0}$, which is additionally damped by $T_{2}$ and the system decays even more rapidly than the normal case. Thus, non-normality contributes to both short-term growth in the system, as well as eventual decay.

\subsection{Finite large scale zonal variability}
\label{Results4}

The $k=0$ limit is useful in understanding the low frequency large scale thermodynamically coupled variability. This section will describe how long but finite large scale variability deviates from this idealized case. As was done for the $k=0$ case (equation \ref{solution}) we can study the stability of the system by constructing a dynamical matrix that contains the WES feedback coupling (\ref{model}) for a general $k$ value and analyze its eigenvalues. The matrix is similar to (\ref{Mk0}), but we use the general functions shown in
\begin{figure}[H]
\centerline{\includegraphics[width=3.125in]{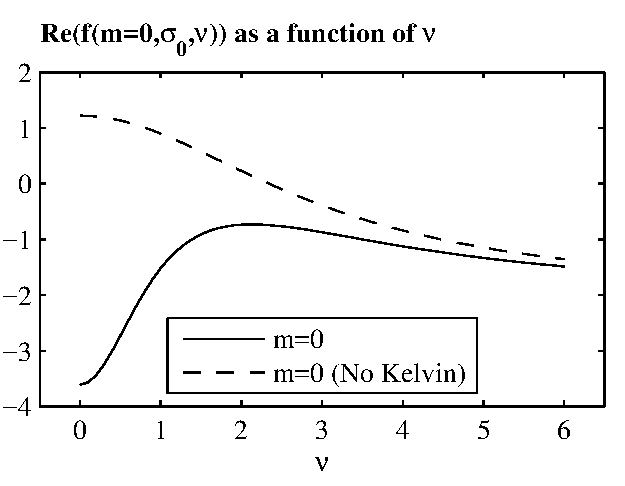}}
\caption{Real part of $f(m=0,\sigma_{0},\nu)$ as a function of $\nu=\frac{k}{\epsilon}$. The solid line contains both the effect of the atmospheric Kelvin and first Rossby waves on $f(m=0,\sigma_{0},\nu)$, while the dashed line contains just the first Rossby wave. Notice that the curves converge as $\nu$ increases, indicating a decreasing influence of the atmospheric Kelvin wave with increasing $\nu$.} \label{f0_NK}   %%%phasing.m
\end{figure}
\noindent (\ref{h_f_functions}). That is
\begin{align}
M_{m,m-2}&=-\frac{1}{2}\epsilon_{T}h(m-1,\nu)\sigma\nonumber\\
M_{m,m}&=\frac{1}{2}f(m,\sigma,\nu)\nonumber\\
M_{m,m+2}&=\frac{1}{2}\epsilon_{T}h(m+1,\nu)\sigma. \label{M}
\end{align}
Note that here we retain the non-normality inherent to the system. Solving equation \ref{solution} for this dynamical matrix, we find that an initial condition $\mathbf{T}(0)=(T_{0}(0),\: T_{1}(0),\:T_{2}(0),\:...)$ evolves as
\begin{equation}
\mathbf{T}(t)=exp(\mathbf{M}t)\mathbf{T}(0)=\mathbf{G}(t)\mathbf{T}(0). \label{G}
\end{equation}
This shows that if all $\mathbf{M}$ eigenvalues $Im(\omega)$ are negative (see equation \ref{solution}) the system is linearly stable and asymptotically decays (although the initial SST anomalies may grow for a finite time due to non-normality as shown in the previous section for $k=0$). If at least one $\omega$ is positive, then the system is linearly unstable, and will be dominated by exponentially growing modes (akin to the normal case shown in figure \ref{non_normal}).

Figure \ref{Im(w)} shows $Im(\omega)$ for the least stable eigenvector of the system as a function of $\nu=\frac{k}{\epsilon}$ for equatorially antisymmetric case, symmetric case, and symmetric with the
atmospheric Kelvin wave suppressed case. For context a zonal wavenumber $k=\frac{2\pi}{120^{o}}$ and $\epsilon=(2d)^{-1}$ corresponds to $\nu=2.44$. For the standard parameters shown in table 1 the system is linearly stable for large zonal scale variability. The system does become unstable for larger $k$ (and stable again for even bigger $k$) but that is well outside the validity of the Gill-Matsuno model under the long wave approximation. The system is stable at all $k$ when the long wave approximation is not used (not shown, see section \ref{Long_wave} for a description of the model used for that effect).

We observe that the stability of the system is greatly reduced for equatorially symmetric variability when we suppress the atmospheric Kelvin wave for very long zonal wavelengths up to $\nu\sim 3.9$. After that point the absence of the Kelvin wave does not seem to affect the stability of the system: note how the dashed and dash-dot curves 
\begin{figure}[H]
\centerline{\includegraphics[width=3.125in]{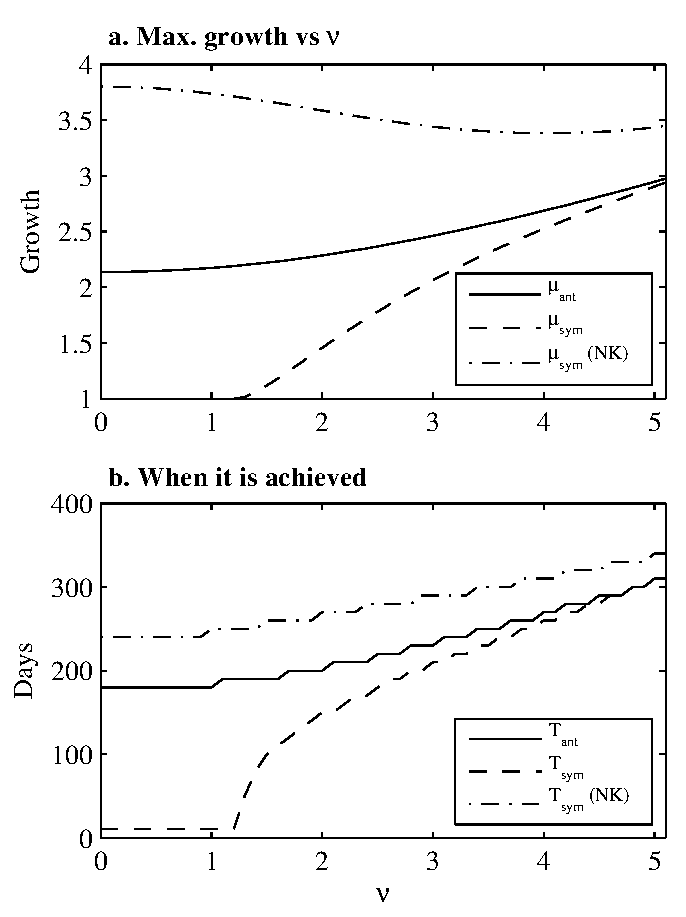}}
\caption{\textbf{(a)} Maximum growth $\mu$ as a function of $\nu$. \textbf{(b)} Time when maximum growth is achieved as a function of $\nu$. The plots cover the linearly stable range under the long wave approximation.} \label{non_normal_growth}
\end{figure}
\noindent converge in figure \ref{Im(w)}b. The atmospheric Kelvin wave gets progressively less important in the dynamics of the system as $\nu$ increases. Figure \ref{f0_NK} shows that the amplitudes of the real part of the growth function $f(m=0,\sigma_{0},\nu)$ with and without the Kelvin wave get closer for increasing $\nu$. Consequently the destructive effect on growth due to the Kelvin wave gets less important as we deviate more and more from the $k=0$ idealized case. This is simply a result of phasing of the Kelvin wave response with respect of the initial forcing: as $k$ gets larger or $\epsilon$ smaller, the Kelvin wave response becomes out of phase with the forcing (see equation \ref{T0_T1}). Both with and without the atmospheric Kelvin wave $f(m=0,\sigma_{0},\nu)$, values should asymptote to $-2$ (see equation \ref{h_f_functions}) as $\nu\rightarrow\infty$. 

Overall, the picture gets much more complicated for non-zero zonal wavenumber. Now the evolution of a particular mode $T_{m}$ will depend on the real and imaginary part of the growth function $f(m,\sigma,\nu)$ as well as in the phasing in the complex plane with modes $T_{m-2}$ and $T_{m+2}$. Nonetheless, the general lessons learned in the $k=0$ case remain valid (albeit much less pronounced) for large scale dynamics. 

Figure \ref{non_normal_growth} shows the maximum transient growth for modes that are equatorially antisymmetric, symmetric, and symmetric with atmospheric Kelvin wave suppressed as a function of $\nu$. Only the linearly stable regime is considered. The framework used to calculate the optimal initial conditions that maximize transient growth in each case is described in appendix IV [see also Farrel and Ioannou (1996)\nocite{Farrell1996}, Vimont (2010)\nocite{Vimont2010}, Martinez-Villalobos and Vimont (2016)\nocite{Martinez2016}]. We observe that the maximum growth for antisymmetric and symmetric variability starts converging as $\nu$ increases. The damping provided by the atmospheric Kelvin wave is still important, but relatively less so as the zonal wavelength gets shorter. This effect is emphasized by comparing the maximum 
\begin{figure}[H]
\centerline{\includegraphics[width=3.125in]{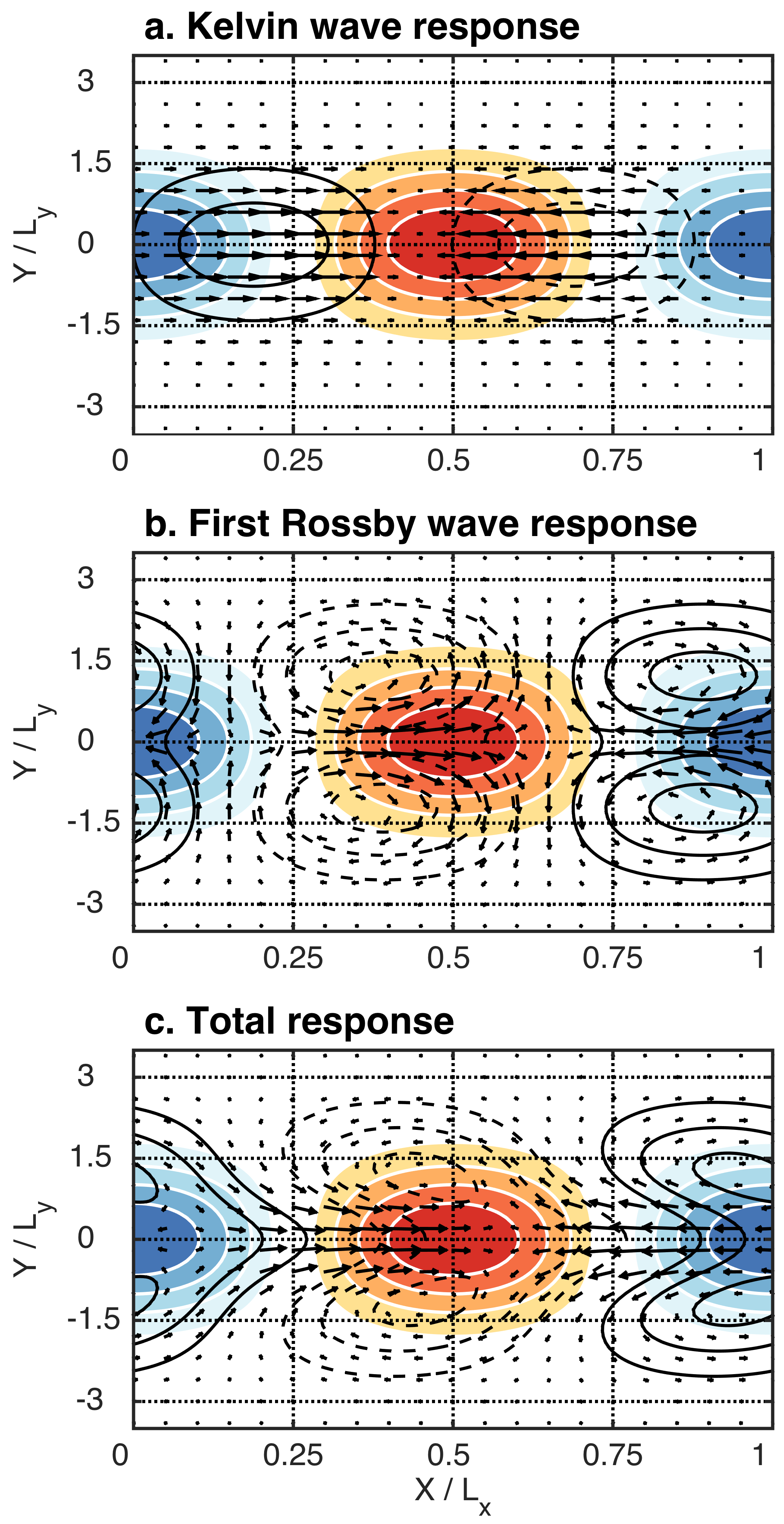}}
\caption{Instantaneous atmospheric wind (vectors) and low level geopotential (contours) response to $SST=\psi_{0}(y)exp(ikx)$. Shown are the: \textbf{(a)} Kelvin wave, \textbf{(b)} first Rossby wave, and \textbf{(c)} total instantaneous atmospheric response. Plots correspond to a zonal wavenumber $L_{x}=120^{o}$ corresponding to $\nu=2.44$, and $L_{y}\sim10^{o}$ (from parameters in Table 1). SST is shaded, and the solid (dashed) contours correspond to positive (negative) low level geopotential anomalies. Units are arbitrary, but consistent between panels.} \label{Instantaneous_sym}
\end{figure}
\noindent growth with and without the Kelvin wave influence in figure \ref{non_normal_growth}.

The reason for the decrease of damping provided by the Kelvin wave as $\nu$ increases, and consequently the increase in similarity between symmetric and antisymmetric structures growth due to the WES feedback, is explained by the relative phasing between the SST and atmospheric response for a finite $\nu$. Figure \ref{Instantaneous_sym} illustrates how the first Rossby wave contributes to growth, and the Kelvin wave contributes to damping for a zonal wavelength of $120^{o}$ ($\nu=2.44$). Notice in figure \ref{non_normal_growth} that for this $\nu$ antisymmetric structures still grow more than symmetric ones, but the gap has decreased tremendously compared to the $k=0$ case. Consequently, we expect a diminished (compared to the $k=0$ case), but still sizable contribution to damping on average by the Kelvin wave.

Figure \ref{Instantaneous_sym}a shows the instantaneous wind anomaly response to a sea surface temperature anomaly pattern of the form $T(x,y)=\psi_{0}(y)exp(ikx)$ associated with the Kelvin wave. Focusing first on the positive SST anomaly located from $x=0.25$ to $x=0.75$ ($x$ in $\frac{2\pi}{k}$ units) in the zonal axis, we notice that over the western part of this pattern (from $x=0.25$ to $x=0.45$) the total wind relaxes (a positive wind anomaly implies a relaxation of the easterly trades), while in the eastern part (from $x=0.45$ to $x=0.75$) the total wind magnitude increases. The relaxation of the wind in the western part implies a decrease in evaporation and a tendency for the SST anomaly to grow there, while in the eastern part the reinforcement of the trades implies a cooling of the warm anomaly. In other words, there is SST growth in the western part and damping in the eastern part. The Kelvin wave acts to damp the anomaly on average because for 
this $\nu$ value the phasing is such that the negative wind anomaly associated with the Kelvin wave overlaps more of the warm anomaly (the eastern part is bigger in extension than the western part). A similar reasoning explains that on average the first Rossby wave acts to grow the anomaly. In this case, the reduced group velocity of the atmospheric Rossby wave means that the wind relaxes over most of the warm SST anomaly (figure \ref{Instantaneous_sym}b). Adding the responses (figure \ref{Instantaneous_sym}c) shows that in this regime the WES feedback acts to grow the initial anomaly. Notice also that this mechanism implies a westward propagation of the structure.

In the $\nu=0$ limit the Kelvin and Rossby wave responses are exactly zonally in-phase with the SST anomalies. In that limit, both negative wind anomalies associated to the Kelvin wave, and positive wind anomalies associated to the first Rossby wave perfectly overlap a SST warm anomaly in the zonal direction, and SST anomalies grow or decay in place with no zonal propagation. For a finite zonal wavelength the phasing of the Kelvin wave changes more rapidly as $\nu$ increases than the phasing induced by the atmospheric Rossby wave, due to the differing group velocities. As a consequence, the Kelvin damping gets reduced more rapidly than the Rossby growth, so on average the symmetric SST mode grows more for shorter, but still long, zonal scales. In summary, in the zonal long wave range, as $\nu$ increases from $0$ the equatorially symmetric mode propagates more efficiently westward achieving more growth in the process compared to smaller $\nu$. 

\begin{figure}[H]
\centerline{\includegraphics[width=3.125in]{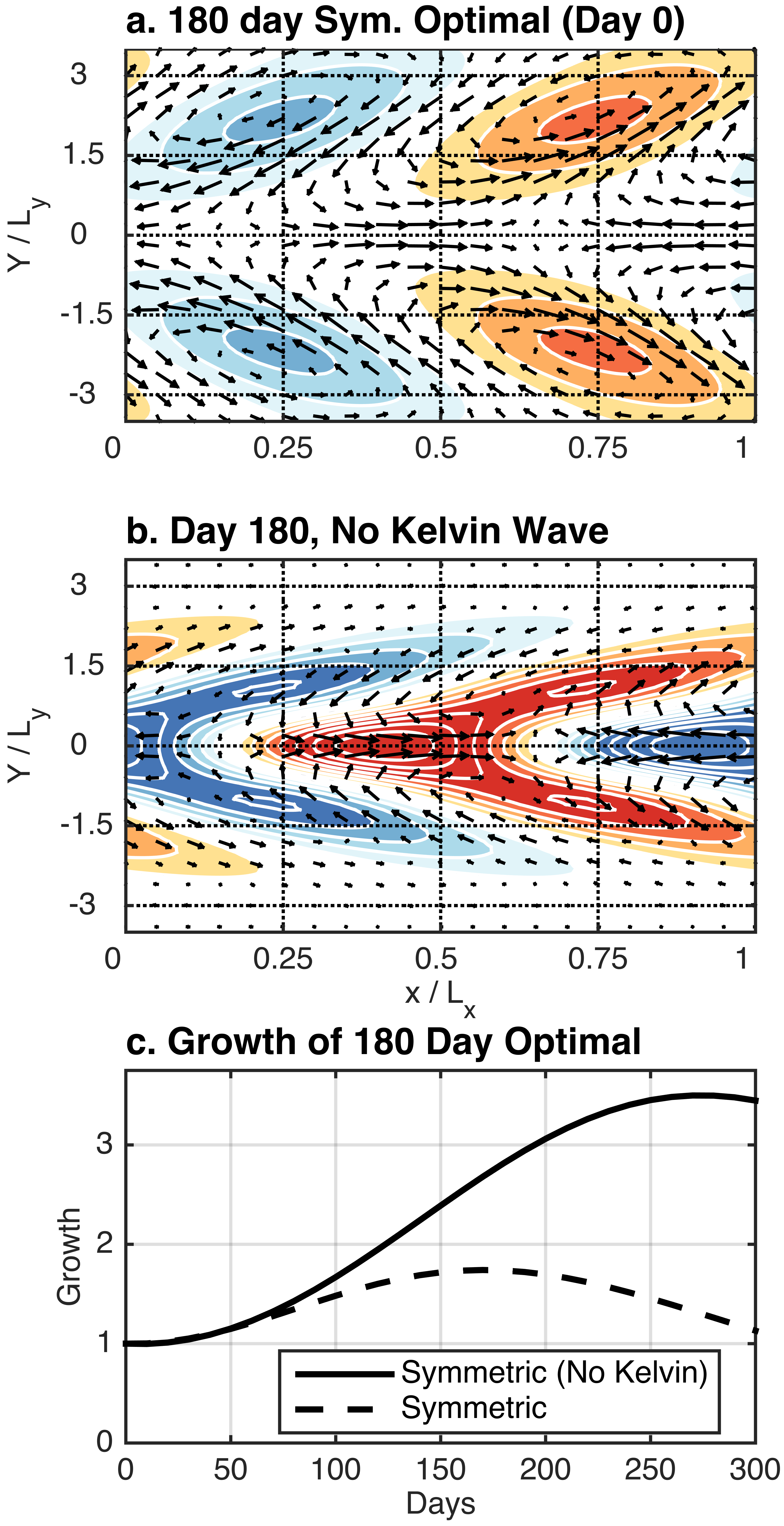}}
\caption{\textbf{(a)} Symmetric ``optimal'' initial condition from Fig.~\ref{Sym_anti_180day}(a). \textbf{(b)} 180 day evolution of the initial condition from \textbf{(a)} for the case when the atmospheric Kelvin wave is suppressed [compare to figure \ref{Sym_anti_180day}(c)]. \textbf{(c)} Growth of the 180 day symmetric optimal with and without the Kelvin wave. For these panels $L_{x}=120^{o}$ $L_{y}$ (the equatorial radius of deformation) $\sim 10^{o}$. Shown is SST (shaded) and low level winds (vectors). Units are arbitrary, but are the same in panels \textbf{(a)}, \textbf{(b)}, and Fig.~\ref{Sym_anti_180day}(a-c).} \label{Sym_NK_180day} %%%From optimal_structures_v.m
\end{figure}

The damping produced by the Kelvin wave is certainly smaller for finite $\nu=\frac{k}{\epsilon}$, but it is still sizable. Figure \ref{Sym_NK_180day}b shows the evolution of the symmetric structure shown in figure \ref{Sym_anti_180day}a with the atmospheric Kelvin wave suppressed. The growth in this case is vastly larger (figure \ref{Sym_NK_180day}c, also compare Fig.~\ref{Sym_NK_180day}b to Fig.~\ref{Sym_anti_180day}c) showing the importance of this mechanism in damping symmetric structures. This explains why equatorially antisymmetric large-scale patterns are preferentially excited by the WES feedback: for symmetric structures, the Kelvin wave plays a critical role in damping variability, while there is no analogous mechanism for anti-symmetric structures.

\subsection{The use of the long wave approximation}
\label{Long_wave}

In any analytical approach some allowance is needed in order to find the adequate balance between analytical understanding and realism. In this study we take an analytical look on the equations that govern thermodynamically coupled variability in hopes of gaining insight on the differences and similarities between equatorially symmetric and antisymmetric modes. To simplify the mathematics and interpretation of the variability we use the long wave approximation. This approximation should give 
\noindent us a qualitative understanding of the variability in the vicinity of the $k=0$ region. Here, we test the validity of the major conclusions away from the $k=0$ limit by relaxing the long wave approximation. 

Vimont 2010\nocite{Vimont2010} and Martinez-Villalobos and Vimont\nocite{Martinez2016} numerically analyze a Gill-Matsuno atmospheric model coupled to a slab ocean without using the long wave approximation under a variety of situations. We will refer to such a model as the ``full model'' as we will use it to contrast the results derived herein using the long wave approximation. In the full model we retain the atmospheric tendencies and add back atmospheric damping in the $v$ equation. [The difference between retaining or dropping 
\begin{figure}[H]
\centerline{\includegraphics[width=3.125in]{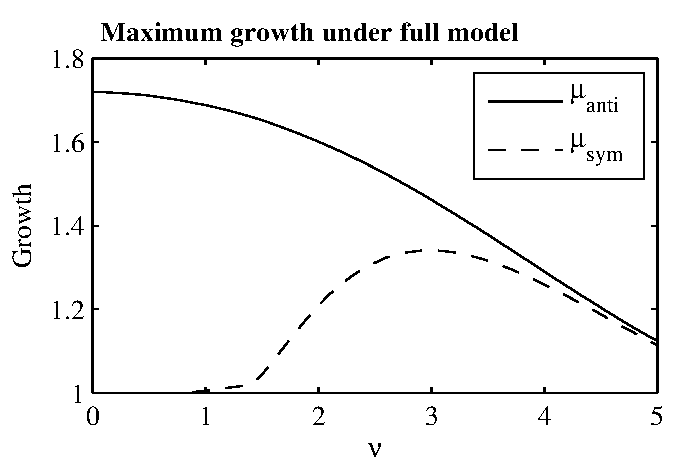}}
\caption{Maximum growth of equatorially symmetric and antisymmetric variability under full model (no long-wave approximation). Compare to figure \ref{non_normal_growth}.}  \label{Full_model}%%%From %long_wave_approx_plots.m
\end{figure}
\noindent the atmospheric tendencies is negligible for the $k$ range we are interested in, but solutions are much more easily computed using equation \ref{G}. See Vimont 2010\nocite{Vimont2010} or Martinez-Villalobos and Vimont 2016\nocite{Martinez2016} for more details.] The resulting equations are projected onto the first 10 parabolic cylinder functions for consistency with the present analysis, and equatorially symmetric and antisymmetric terms are collected. The resulting equations may be written in the form of equation \ref{solution} with solution shown in (\ref{G}), but in this case the state vector also contains the atmospheric variables. 

Figure \ref{Full_model} shows the maximum SST growth of the first symmetric and antisymmetric optimal as a function of $\nu$ under the full model. We compare this figure with figure \ref{non_normal_growth}a that shows the same curves (symmetric and antisymmetric) but under the long wave approximation. At first look the symmetric and antisymmetric maximum growth in both models don't look similar, but a closer look reveals more similarities than differences. The symmetric optimal does not experience growth for small $k$, and starts to grow roughly at the same value of $\nu\sim 1$. The antisymmetric optimal maximum growth is always bigger than the maximum symmetric growth (for this parameter regime), and both maximum symmetric and antisymmetric growth approach each other as $k$ increases, likely because the Kelvin wave also kills growth in the full model and its effect is reduced for shorter zonal scales.

The symmetric variability is very well approximated at small $k$ while there is an error in the antisymmetric variability even at small $k$ (the $\epsilon v$ term dominates the meridional momentum equation very close to the equator for antisymmetric variability at any $k$, while for the symmetric mode $\epsilon v$ is strictly zero at the equator and small in the vicinity of the equator). In many respects the long wave approximated model acts like a less damped version of the full model, which explains the smaller growth shown in general in Fig.~\ref{Full_model} compared to figure \ref{non_normal_growth}a. That parallel is not exact for the antisymmetric variability. In the full model the maximum growth occurs at $k=0$ in agreement with Xie et al. 1999 \nocite{Xie1999} while in the long wave approximation the maximum growth occurs for $\nu>0$. In spite of these differences the spatial structures of the symmetric and antisymmetric optimals (and also the time evolved structures) for the full and long wave approximated model look almost identical, showing that the long wave approximation is a good representation of the WES feedback process, despite differences in the maximum growth attained (less damped than the full model).

\section{Conclusions and final comments}
\label{Concl}

\subsection{Conclusions}
\label{Concl_sec1}

The growth and equatorial symmetry properties of thermodynamically coupled ocean-atmosphere variability was studied using the Gill-Matsuno atmospheric model coupled to a thermodynamic slab ocean. Assuming an atmosphere completely determined by the underlying SSTs the ocean and atmosphere variables were projected onto parabolic cylinder functions $\psi_{m}(y)$, effectively decomposing the variability onto different $T_{m}$ modes. Under the assumption of a geographically homogeneous coupling (i.e. $K_{q}$ and $\alpha$ constant) two independent sets of solutions emerge: equatorially symmetric ($m$ even), and equatorially antisymmetric (meridional mode-like, $m$ odd). These two sets of solution behave similarly away from the equator, for $T_{m}$ modes $m\geq2$, and differ significantly in the equatorial zone, for $T_{m}$ modes $m=0,1$. The main difference in the equatorial zone is traced to the additional SST damping provided by the atmospheric Kelvin wave for equatorially symmetric variability for large zonal scales. On the other hand antisymmetric variability close to the equator is unaffected by a damping term analogous to the Kelvin one for symmetric variability, leading to more growth on average. 

It is found that both equatorially symmetric and antisymmetric variability propagates in a similar manner towards the equator. In this framework this is realized by a decrease in amplitude of higher $T_{m}$ modes and an increase in amplitude of lower $T_{m}$ modes. When the variability reaches the equatorial zone it excites the $m=0$ mode in the symmetric case, and the $m=1$ mode in the antisymmetric case, leading to two very different outcomes. In the symmetric case the SST signal is subjected to additional damping by the atmospheric Kelvin wave, while in the antisymmetric case the SST signal grows through the positive WES feedback produced by the first antisymmetric atmospheric Rossby wave. For large zonal scales the outcome of this process is growth of the antisymmetric mode, and decay of the symmetric mode. In other words any SST distribution will grow effectively through this mechanism if its meridional structure has a large projection onto the $\psi_{1}(y)$ function (equatorially antisymmetric SST structure peaking around $8^{o}$ of latitude for the parameters in this study, also see figure \ref{Hermite}). On the other hand an equatorially symmetric SST distribution confined to the equator [large projection onto the $\psi_{0}(y)$ function] will be short lived in absence of other feedbacks. For shorter zonal wavenumber, but still relatively large zonal scale, damping of symmetric structures by the Kelvin wave is reduced and the growth of both symmetric and antisymmetric variability is similar. 

Under the range of parameters considered in this study the growth process of thermodynamically coupled variability is fundamentally non-normal. This non-normality is manifest in interactions that initially grow a SST anomaly while at the same time seeding its eventual decay. Initial growth occurs as a given SST structure (without loss of generality, we consider an initial positive anomaly) produces a positive tendency for lower-order modes, and negative tendency for higher order modes. These tendencies lead to short-term growth. The positive tendency of the lower-order mode excites the next-lowest-order mode, and so forth until the lowest order anti-symmetric ($m=1$) or symmetric ($m=0$) structure is excited, at which that mode either grows on its own (for the $m=1$ anti-symmetric structure) or experiences enhanced decay due to the Kelvin wave (for the $m=0$ symmetric structure, as described above). At the same time (again assuming initial positive polarity for a given mode), as the amplitude of the higher order mode grows more large and negative, the negative amplitude of the higher order mode in turn contributes to decay of the original SST mode. The non-normal decay process eventually counteracts growth of the total anomaly, and is critical for maintaining stability of the system (under the parameter regime considered herein). In principle this growth process does not require equatorial antisymmetry and in fact it would prefer the symmetric structure should the Kelvin wave not exist.

Using this framework we show that both equatorially symmetric and antisymmetric modes propagate westward as a consequence of the group velocity of atmospheric Rossby waves. In the antisymmetric case the propagation is due to the zonal phasing between the antisymmetric Rossby wave wind response and SST anomalies as found by Xie et al. 1999 \nocite{Xie1999a}. In the symmetric case, the interpretation is the same for $m>2$ symmetric Rossby waves, except in the equatorial zone. In the equatorial region the propagation direction will depend on the relative phasing between the first atmospheric Rossby and Kelvin waves. It turns out that for large zonal scales, this phasing favors the growth by the first Rossby wave implying a westward propagation of the SST structure.

\subsection{Discussion}
\label{Concl_sec2}

The implications of this study, although assuming a homogeneous coupling, can be extended into a system that includes geographical variations in the mean state, as manifest by a geographic dependence of $K_q$ \cite{Martinez2016}. In that case, atmospheric heating depends on the product $K_q(y) T$, and hence the mean state acts as a ``mode selector'' for enhancing coupling of particular modes. As an example consider the air-sea coupling distribution provided by an equatorially symmetric but meridionally thin Inter Tropical Convergence Zone (ITCZ). This coupling structure will suppress off-equatorial modes and enhance just the atmospheric Kelvin and lowest-order Rossby waves. In the limit of a vanishingly thin ITCZ (in which $K_q(y)$ approaches a delta-function), coupling for antisymmetric modes are eliminated (via the product of $K_q(y)$ and antisymmetric $T_{m}$) and we would expect just a thermodynamically coupled symmetric mode to be able to grow, as shown in Martinez-Villalobos and Vimont (2016)\nocite{Martinez2016}. Following the same idea, a broader symmetric ITCZ would produce a situation more akin to the one presented in this paper: more growth for antisymmetric modes, because the symmetric modes are additionally damped by the Kelvin wave. This interpretation could be extended to include variations in the assumed atmospheric radius of deformation as well, e.g. the case of a Battisti 1999\nocite{Battisti1999b} or Lindzen and Nigam 1987\nocite{Lindzen1987} style model of the atmospheric boundary layer. Similar arguments could also be applied to understanding zonal evolution through a zonally and meridionally varying mean state.

This framework also helps understand the consequences of equatorial asymmetry in the coupling, for example through meridional asymmetries in the ITCZ position. As discussed above, a narrowly confined equatorially symmetric ITCZ (e.g. tropical Atlantic during boreal spring) suppresses coupling for off-equatorial and anti-symmetric modes and enhances the atmospheric Kelvin and first symmetric Rossby responses. In contrast, for a narrow ITCZ placed off the equator (e.g. tropical Atlantic and eastern Pacific during boreal fall) the coupling is asymmetric. In that case, preference for equatorially symmetric or antisymmetric modes will depend on the relative projections of the $\psi_{1}(y)$ and $\psi_{0}(y)$ structures onto $K_q(y)$. For a narrow ITCZ structure centered near a maxima in $\psi_1(y)$, antisymmetric modes will dominate the solution (even with a non-zero projection of $K_q(y)$ onto $\psi_0(y)$) due to the damping effect of the Kelvin wave for the symmetric component of the solution. In that case however, the solution will include a mix of both symmetric and antisymmetric components. This mechanism explains the findings of Martinez-Villalobos and Vimont 2016\nocite{Martinez2016} in that regard (compare figures 4 and 5 in that paper). This interpretation may also be useful for interpreting similar variability in models with less restrictive coupling parameterizations, e.g. \cite{Lindzen1987}, \cite{Battisti1999b}.

This work shows that both anti-symmetric and symmetric structures emerge from the physical processes that generate tropical ``meridional modes''; i.e. both anti-symmetric and symmetric structures grow and propagate due to the same physical processes. This is somewhat in contrast to early discussions of meridional mode behavior, where it was thought that equatorial anti-symmetry would enhance growth through providing an interhemispheric temperature gradient that would enhance the surface wind response. Given the results herein, it appears that an interhemispheric gradient is not necessary, and in fact equatorially symmetric structures would preferentially grow in the absence of the Kelvin wave. Instead, growth is more closely related to the zonal and meridional phasing between the atmospheric Rossby wave response to imposed heating and the SST that is presumably responsible for that heating in the first place \cite{Vimont2010}. Based on these arguments, one might ask whether the term ``meridional mode'' is still appropriate. We show herein that both anti-symmetric and symmetric structures exist as a class of structures that evolve in a similar manner towards lower meridional wavenumbers; as such, we argue that the phrase ``meridional mode'' is appropriate for both anti-symmetric and symmetric structures.

The aim of this study is to explain the main similarities and differences between symmetric and antisymmetric thermodynamically coupled variability using a common framework. As such, besides the usual linearity assumption, there are many approximations being made and a lot of room to improve the model. For example the parameters in the model, especially the coupling, do not vary geographically. There are different regions in the tropical oceans where we would not expect the same WES feedback coupling. Another important approximation is the assumption that latent heat is unaffected by meridional wind variations. These variations will affect antisymmetric and symmetric modes differently, and we hope to consider them in a subsequent study. (The effect of mean meridional winds has been investigated by Liu and Xie 1994\nocite{Liu1994}, and Wang 2010\nocite{Wang2010}). Also, the use of the long wave approximation produces less damped modes than using a model with meridional wind damping, although that does not affect the spatial structure of the modes. Finally variations of other fluxes, specially short wave flux, should be considered in the future. Despite these caveats, this study does provide physical insight into the workings of thermodynamically coupled variability. The basic meridional mode coupling mechanism is contained within the framework here presented. We expect that the insight achieved here, by stripping down the WES interaction into its most basic building block, can be translated into explaining observed variability and more complex model output, especially slab ocean simulations.

In nature, equatorially antisymmetric modes coupled through the WES feedback are an established and relevant part of our climate system, and are the subject of a large literature collection studying meridional mode variability in both the tropical Atlantic and Pacific. On the other hand, thermodynamic equatorially symmetric variability has been less studied, likely because the Bjerknes feedback masks these kinds of modes. Nonetheless the interest in this kind of variability has increased following the findings of Clement et al. 2011\nocite{Clement2011}. In addition the Pacific Meridional Mode has an important degree of equatorial symmetry \cite{Chiang2004}, and as such the results derived herein may be relevant in nature. We hope that this study provides a common ground to understand these modes in a unified framework.

\section*{acknowledgments}
This work was supported by NSF
Climate and Large Scale Dynamics Projects ATM-
0849689 and 1463970 and the University of Wisconsin
Climate, People, and the Environment Program.

\appendix
\label{appendix}

\section*{Appendix I: Parabolic cylinder functions}
\label{appendix_hermite}

The version of the parabolic cylinder functions used in this study are given by:
\begin{equation}
\psi_{m}(y)=\frac{1}{\sqrt{2m!\pi^{1/2}}}H_{m}exp(-y^2/2)
\end{equation}
where $H_{m}$ are the physicist Hermite polynomials. These polynomials are defined for non negative $m$ integers. The first 4 functions are plotted in Fig.~\ref{Hermite}.

The functions are orthonormal
\begin{equation}
\int_{-\infty}^{\infty}\psi_{m}(y)\psi_{n}(y)dy=\delta_{mn}. \label{normalization}
\end{equation}

\section*{Appendix II: Atmospheric variables in terms of SSTs}
\label{appendix_atmosphere}

We write the atmospheric variables using the parabolic cylinder functions as
\begin{align}
u(t,x,y)&=\sum_{m=0}^{\infty}u_{m}(t)\psi_{m}(y)exp(ikx) \nonumber \\
v(t,x,y)&=\sum_{m=0}^{\infty}v_{m}(t)\psi_{m}(y)exp(ikx) \nonumber \\
\phi (t,x,y)&=\sum_{m=0}^{\infty}\phi_{m}(t)\psi_{m}(y)exp(ikx) \label{atmosphere}
\end{align}
In solving the model equations for a particular mode $m$ the following identity is useful:
\begin{align}
y\psi_{m}&=\sqrt{\frac{m+1}{2}}\psi_{m+1}+\sqrt{\frac{m}{2}}\psi_{m-1} \nonumber \\
\frac{d\psi_{m}}{dy}&=-\sqrt{\frac{m+1}{2}}\psi_{m+1}+\sqrt{\frac{n}{2}}\psi_{m-1}, \label{identity}
\end{align}
as well as defining the auxiliary variables \cite{Gill1980}:
\begin{align}
q=\phi+u\nonumber\\ 
r=\phi-u \label{auxiliary}
\end{align}

Plugging equation \ref{atmosphere} into the model equations (\ref{model}), and using equations \ref{decomposition}, \ref{identity} and \ref{auxiliary} we find that
\begin{align}
q_{0}&=-K_{q}\frac{T_{0}}{\epsilon+ik}\nonumber\\
q_{m+1}&=K_{q}(\frac{\sqrt{m(m+1)}T_{m-1}+mT_{m+1}}{-(2m+1)\epsilon+ik}). \label{q}
\end{align}
In doing so we used that
\begin{equation}
r_{m-1}=\sqrt{\frac{m+1}{m}}q_{m+1}, \label{qr}
\end{equation}
which is a consequence of making the long-wave approximation \cite{Gill1980} in the meridional wind equation.
Inverting equation \ref{auxiliary} and using equation \ref{qr} we get 
\begin{align}
u(t,y,x)&=\frac{1}{2}(q_{0}(t)\psi_{0}(y)+\sum_{m=1}^{\infty}q_{m+1}(t)(\psi_{m+1}(y)-\sqrt{\frac{m+1}{m}}\psi_{m-1}(y)))exp(ikx)\nonumber\\
\phi(t,y,x)&=\frac{1}{2}(q_{0}(t)\psi_{0}(y)+\sum_{m=1}^{\infty}q_{m+1}(t)(\psi_{m+1}(y)+\sqrt{\frac{m+1}{m}}\psi_{m-1}(y)))exp(ikx).\label{uphi}
\end{align}
\renewcommand{\thefigure}{A\arabic{figure}} %This is to st
\setcounter{figure}{0}
\begin{figure}[H]
\centerline{\includegraphics[width=3.125in]{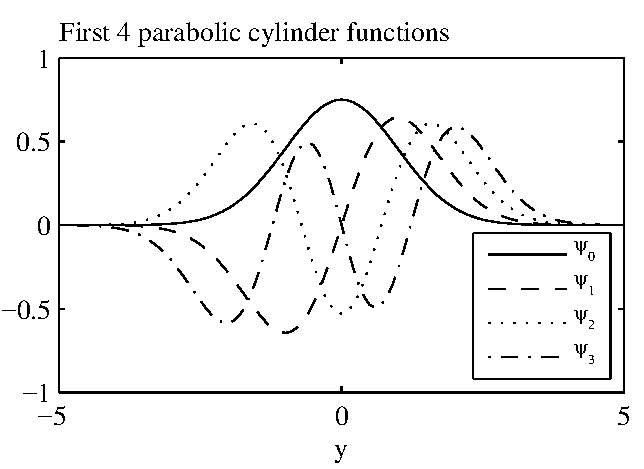}}
\caption{First four parabolic cylinder functions.} \label{Hermite}%%%From optimal_structures.m
\end{figure}
\noindent The meridional wind can be found by plugging the above equations into the zonal wind and geopotential equations and solving for $v$ yielding:
\begin{equation}
v(t,y,x)=\sum_{m=0}^{\infty}\sqrt{\frac{1}{2(m+1)}}(K_{q}T_{m+1}(t)+(\epsilon+ik)q_{m+1}(t))\psi_{m}(y)exp(ikx)\label{v}
\end{equation}
Equations \ref{uphi} and \ref{v} correspond to equation \ref{atm} in the main text.

\section*{Appendix III: Growth equation}
\label{appendix_growth}

Using equations (\ref{decomposition}) and (\ref{atmosphere}) we can write the equation for the evolution of $T_{m}$ and $T_{m}^{*}$ as
\begin{align}
\frac{\partial T_{m}}{\partial t}&=\alpha u_{m}-\epsilon_{T}T_{m}\nonumber\\
\frac{\partial T^{*}_{m}}{\partial t}&=\alpha u^{*}_{m}-\epsilon_{T}T^{*}_{m}
\end{align} 
Multiplying the first equation by $T^{*}_{m}$ and the second one by $T_{m}$ and adding them we get
\begin{equation}
\frac{\partial |T_{m}|^2}{\partial t}=\alpha(u_{m}T^{*}_{m}+u^{*}_{m}T_{m})-2\epsilon|T_{m}|^2.
\end{equation}
Using equations \ref{auxiliary}, \ref{qr} and \ref{q} we find that the growth of a particular mode is governed by

\begin{align}
\frac{\partial |T_{m}|^2}{\partial t}&=K_{q}\alpha[-(\frac{1}{((2m-1)^2\epsilon^2+k^2)})((m-1)(2m-1)\epsilon)|T_{m}|^2  \nonumber \\
&+(\frac{1}{((2m+3)^2\epsilon^2+k^2)})((m+2)(2m+3)\epsilon)|T_{m}|^2 \nonumber \\
&-(\frac{1}{((2m-1)^2\epsilon^2+k^2)})(\sqrt{m(m-1)}(2m-1)\epsilon)|T_{m}||T_{m-2}|cos(\delta_{m}-\delta_{m-2}) \nonumber \\
&+(\frac{1}{((2m+3)^2\epsilon^2+k^2)})(\sqrt{(m+2)(m+1)}(2m+3)\epsilon)|T_{m}||T_{m+2}|cos(\delta_{m}-\delta_{m+2}) \nonumber \\
&-(\frac{1}{((2m-1)^2\epsilon^2+k^2)})(\sqrt{m(m-1)}k)|T_{m}||T_{m-2}|sin(\delta_{m}-\delta_{m-2}) \nonumber \\
&+(\frac{1}{((2m+3)^2\epsilon^2+k^2)})(\sqrt{(m+2)(m+1)}k)|T_{m}||T_{m+2}|sin(\delta_{m}-\delta_{m+2})]-2\epsilon_{T}|T_{m}|^2. \label{growth_eqn}
\end{align}
Here $\delta_{m}-\delta_{m-2}$ is the phasing in the complex plane between modes $m$ and $m-2$, and comes from the decomposition $T_{m}=|T_{m}|exp(i\delta_{m})$.

To find the equation for total SST growth we use equation \ref{decomposition} and write the total SST amplitude squared $|T|^2$ in terms of the individual modes.
\begin{equation}
|T|^2=\sum_{m=0}^{\infty}\sum_{n=0}^{\infty}T_{m}^{*}T_{n}\psi_{m}\psi_{n}
\end{equation}
Integrating $|T|^2$ over the whole domain (the symbol $<>$ denotes that integration) we get the total growth that can be decomposed in terms of the growth of individual terms as follows
\begin{equation}
<|T|^2>=\int_{-\infty}^{\infty}\int_{0}^{L_{x}}|T|^2dxdy=L_{x}\sum_{m=0}^{\infty}|T_{m}|^2, \label{Tm2}
\end{equation}
where $L_{x}$ is the zonal extension of the basin considered and we have used equation \ref{normalization}.
Using equation \ref{growth_eqn} the equation for the total SST growth is given by
\begin{align}
\frac{\partial}{\partial t}<|T|^2>=&L_{x}K_{q}\alpha\sum_{m=0}^{\infty}[-(\frac{1}{((2m-1)^2\epsilon^2+k^2)})((m-1)(2m-1)\epsilon)|T_{m}|^2  \nonumber \\
&+(\frac{1}{((2m+3)^2\epsilon^2+k^2)})((m+2)(2m+3)\epsilon)|T_{m}|^2 \nonumber \\
&+(\frac{2}{((2m+3)^2\epsilon^2+k^2)})(\sqrt{(m+2)(m+1)}k)|T_{m}||T_{m+2}|sin(\delta_{m}-\delta_{m+2})]\nonumber \\
&-2L_{x}\epsilon_{T}\sum_{m=0}^{\infty}|T_{m}|^2.
\end{align} 
Notice there is a partial cancellation between the exchange terms.

\section*{Appendix IV: Growth optimization}
\label{appendix_optimization}

To find the initial conditions that maximize SST growth by the WES feedback at some time later $\tau$ we use a standard Lagrangian multiplier approach. The function $\mathcal{L}_{\tau}$ contains the information to optimize.
\begin{equation}
\mathcal{L}_{\tau}=|\mathbf{T}(\tau)|^2-\lambda (|\mathbf{T}(0)|^2-c).
\end{equation}
Here $\mathbf{T}(\tau)$ corresponds to the vector $(T_{0}(\tau),\: T_{1}(\tau),\:T_{2}(\tau),...)$. The first term in the right hand side $|\mathbf{T}(\tau)|^2=\sum_{m=0}^{\infty}|T_{m}(\tau)|^2$ (compare to equation \ref{Tm2}, we are optimizing $\frac{1}{L_{x}}<|T(\tau)|^2>$) is the quantity to be optimized, while the second term is the constraint, i.e. initial condition amplitude equal to some arbitrary constant $c$. Using (\ref{G}) and taking the derivative of the previous equation respect to $T(0)$, we find that the initial condition $T(0)$ that maximizes growth at time $\tau$ satisfies the following eigenvalue problem:
\begin{equation}
\mathbf{G^{\dagger}}(\tau)\mathbf{G}(\tau) \mathbf{T}(0)=\lambda \mathbf{T}(0),
\end{equation}
where $\lambda$ corresponds to the growth at time $\tau$.

%\bibliography{Bib_arxiv_2}  %This is where I have those files located
\bibliography{Paper_arxiv2.bib}
\bibliographystyle{apalike} %This is where I have those files located

% * <cmartinezvil@wisc.edu> 2016-07-25T17:30:35.230Z:
%
% ^.
\end{document}